\documentclass[12pt, a4paper]{article}

\usepackage[utf8]{inputenc}
\usepackage[T1]{fontenc}
\usepackage[margin=1in]{geometry}
\usepackage{setspace}
\usepackage{authblk} 

\usepackage[numbers,square]{natbib} 

\usepackage{amsmath, amssymb, amsfonts, bm, upgreek, mathrsfs, gensymb}

\usepackage{graphicx, booktabs, dcolumn, placeins, xcolor}

\usepackage{algorithmic, textcomp}

\newcommand{\beq}{\begin{equation}}
\newcommand{\eeq}{\end{equation}}
\newcommand{\bdis}{\begin{displaymath}}
\newcommand{\edis}{\end{displaymath}}
\newcommand{\bea}{\begin{eqnarray}}
\newcommand{\eea}{\end{eqnarray}}
\newcommand{\barr}{\begin{array}}
\newcommand{\earr}{\end{array}}
\newcommand{\bfig}{\begin{figure}[!h]}
\newcommand{\efig}{\end{figure}}

\title{\textbf{Modeling of a magnetic field sensor based on spin Hall magnetoresistance}}

\author[1,4]{Syeda Farwa Bukhari}
\author[1]{Alessandro Magni\thanks{Corresponding author: a.magni@inrim.it}}
\author[2]{Witold Skowro\'{n}ski}
\author[1]{Elena Losero}
\author[1]{Vittorio Basso}
\author[1]{Carlo Appino}
\author[2]{Piotr Wi\'{s}niowski}
\author[3]{Juergen Langer}
\author[3]{Berthold Ocker}
\author[4]{Dario Daghero}
\author[1]{Michaela Kuepferling\thanks{Corresponding author: m.kuepferling@inrim.it}}

\affil[1]{Istituto Nazionale di Ricerca Metrologica, Torino, Italia}
\affil[2]{AGH University of Krakow, Institute of Electronics, Krakow, Poland}
\affil[3]{Singulus Technologies AG, Kahl am Main, Germany}
\affil[4]{DISAT, Politecnico di Torino, Torino, Italy}

\date{}

\begin{document}

\maketitle

\begin{abstract}
Next-generation spintronic sensors aim to overcome the limitations of traditional tunneling-magnetoresistance (TMR) devices, such as complex manufacturing, high $1/f$ noise, and significant offsets. This work presents a comprehensive modeling and experimental validation of a magnetic field sensor based on Spin Hall Magnetoresistance (SMR) in a Wheatstone bridge configuration. Utilizing a multiphysics approach, we simulate the interplay between SMR, Anisotropic Magnetoresistance (AMR), and Spin-Orbit Torque (SOT) using a Stoner-Wohlfarth model complemented by a Fuchs-Sondheimer analysis of current distribution. To account for the presence of magnetic domains, we incorporate a modified Stoner-Wohlfarth framework that considers non-uniform magnetization and domain wall motion through a "truncated astroid" approach, allowing for a statistical distribution of single-domain particles. The model is validated against experimental measurements of Pt/$\text{Fe}_{60}\text{Co}_{20}\text{B}_{20}$ and Ta/$\text{Fe}_{60}\text{Co}_{20}\text{B}_{20}$ bilayers patterned into Hall bars and Wheatstone bridges. The model provides critical design guidelines for optimizing material properties, layer thickness, and device layout to minimize power consumption and maximize sensitivity in SMR-based sensing applications.

\vspace{1em} 
\noindent \textbf{Keywords:} Anisotropic magnetoresistance, Spin Hall magnetoresistance, spin orbit coupling, spin orbit torque, Stoner-Wohlfarth

\end{abstract}

\section{Introduction}
Magnetic field sensors are used in a number of applications, spanning from data  storage (sensing the stray fields of the magnetized bits on hard disk drives) and automotive systems, where various currents and angular velocities are measured indirectly \cite{Dieny-2020}, to biomedical applications, in which  the magnetic fields generated by the heart or brain are monitored for diagnosis \cite{Chopin-sensor-2020, Kanno2022}. Sensors based on the Hall effect, which generate a  transverse voltage linearly proportional to the current and magnetic field, are  the most widely employed commercially \cite{Tumanski2011}. Additionally, various types of magnetoresistive sensors exist\cite{Ripka-2010}, taking advantage of the resistance dependence on the relative orientation of the magnetization and current (anisotropic magnetoresistance - AMR) or two magnetic layers (giant- or tunneling-magnetoresistance - GMR, TMR) \cite{Freitas2016,
Zheng_sensors_2019, khan2021magnetic, leitao2024enhanced}. The latter are based on spin-polarized currents, and their response depends on the angle between the magnetization vectors in two separated magnetic layers.

These novel spintronic magnetic field sensors are increasingly replacing traditional magnetic field sensors, such as Hall sensors, because they offer higher sensitivity and better miniaturization perspectives \cite{Oogane_2021, Byeonghwa_2022}. However, the most prominent spintronic sensor, which exploits tunneling magnetoresistance, still suffers from a series of drawbacks that need to be addressed in order to reach a wider market. Among these are the complex manufacturing of reproducible devices, the presence of large 1/f noise, and offset.

To address these limitations, next-generation spintronic sensors are being developed\cite{yihong2021}. A promising route is to exploit the Spin Orbit Torque (SOT) in bilayers composed of metals with Spin Orbit coupling contributions and ferro-or ferrimagnetic materials \cite{Miron-2011, li2021spin, Miron_2010} to improve stability. It further provides effective methods for offset compensation and resetting, similar to the flipping coils in AMR sensors \cite{Korlatan_SOT_2023,suess2021}. Another way to innovate spintronic sensors was proposed recently by exploiting the spin-Hall magnetoresistance (SMR), which takes advantage of the spin current generation in materials with spin-orbit coupling, such as heavy metals: Ta, W, Pt, and its torque on the adjacent ferromagnet \cite{Xu-SOT_2019}. It was demonstrated that the SMR signal can effectively sense magnetic fields, achieving a detectivity of $1 \text{ nT/Hz}^{1/2}$, while maintaining low power consumption and a simple device structure \cite{Xu-SOT_2019, Xu2017}. The simple structure, consisting of a thin film bilayer of a few nanometers, allows for multifunctional applications where semi-transparency for optical access is required \cite{Yang2017}.

The operation and performance of such a sensor are driven by an interplay between SMR, AMR, and SOT. Additionally, weak magnetocrystalline anisotropy and the Oersted field may contribute to the sensor output. The work aims to model the behavior of an SMR based sensor and to validate the model  against experimental results, thus providing indications on how to optimize the performance based on material properties and design, in particular, to reduce its power consumption.

A multiphysics approach was used to model the current distribution, magnetoresistance and magnetic hysteresis of the multilayer structures. The current density distribution in the multilayer is analyzed using a Fuchs-Sondheimer approach \cite{fuchs1938conductivity}, from which the magnetoresistance contributions can be derived considering a parallel connection of two resistive layers with distinct magnetoresistive effects, namely AMR and SMR. SOT is taken into account through the Landau-Lifshitz-Gilbert model, with parameters obtained by the spin diffusion model and a thermodynamical approach \cite{Liu-2020,Magni2022}. The hysteretic dependence of the resistance on magnetization is modeled with a Stoner-Wohlfarth-based approach. In particular, the correlation between resistance as a function of field and domain distribution and size is taken into account \cite{Appino-2000}, an aspect that is little considered in the literature. 

To validate the model, simple Wheatstone bridge type structures fabricated from heavy metal (HM)/ ferromagnetic material (FM) bilayers (where HM is chosen from Pt, Ta, and FM is Fe$_\mathrm{60}$Co$_\mathrm{20}$B$_\mathrm{20}$) and measured. A preliminary characterization of the SOT efficiency parameters was performed by applying the spin diffusion model \cite{Liu-2020,Magni2022} to measurements obtained from  Hall bar devices with varying  HM thicknesses.   

\section{Spin-orbit-torque sensors} \label{sect:SOT}

While traditional Giant Magnetoresistance (GMR) and Tunnel Magnetoresistance (TMR) sensors require at least two ferromagnetic layers, Spin-Orbit Torque (SOT) enables the replacement of the spin-polarizing layer with a non-magnetic conductor. This approach can improve immunity to stray fields \cite{An2025, Rohrmann2018} and offers efficient methods for linearizing sensor output or eliminating the need for flipping coils \cite{suess2021}.

The SOT is the result of an effective field due to a spin current $J_s$ generated by either the spin Hall effect (SHE) or the Rashba-Edelstein effect~\cite{Nguyen2021}. While the first is considered a bulk effect (although it might have an interfacial component) \cite{Hoffmann2013} the second is a pure interfacial effect often present in 2D materials or oxide interfaces \cite{Manchon2015}. SOT is driving magnetization dynamics and altering the magnetization ($M_\mathrm{S}$) configuration. In a macrospin approach (assuming the magnetization is constant with only its direction changing), the SOT can be described within the torque equation. 

The torque equation, or stationary Landau-Lifshitz-Gilbert (LLG) equation \cite{Bertotti-2009}, is given by:
\begin{equation}
    \frac{d \mathbf{M}}{dt}- \frac {\alpha_{\mathrm{d}}} {M_S} \mathbf{M}\times \frac{d \mathbf{M}}{dt} = -\mu_0 \gamma_G \mathbf{M} \times \mathbf{H}_{\mathrm{eff}} + \mathbf{T}_{\mathrm{SOT}}
    \label{Eq:LLG}
\end{equation}
where $\mathbf H_{\mathrm{eff}}$ is the effective magnetic field resulting from the different magnetic energy terms, $\alpha_{\mathrm{d}}$ is the damping constant, and $\gamma_G$ is the gyromagnetic ratio and $\mu_\mathrm{0}$ is the vacuum permeability. The spin-orbit-torque term includes both the field-like and the damping-like terms \cite{Haney-2013}:
\begin{equation}
    \mathbf{T}_{\mathrm{SOT}}= \frac{\mu_B/e}{d_{FM}} j_e [ \xi_{FL} \mathbf{m} \times \mathbf{s} + \xi_{DL} \mathbf{m} \times (\mathbf{m} \times \mathbf{s})]
    \label{Eq:TST}
\end{equation}
where $\xi_{DL}$ and $\xi_{FL}$ are the damping and field like torque efficiencies and $\mathbf{s}$ is the polarization direction of the injected spin current, $\mu_B$ is the Bohr magneton, $e$ is the electron charge, $d_{FM}$ is the thickness of the ferromagnetic film, and $j_\mathrm{e}$ is the charge current density.

The effect of SOT on the magnetization can be described by a damping-like and field-like torque field $\mathbf H_{DL}$, $\mathbf H_{FL}$, which can be obtained by \cite{Magni2022, Basso2022}:
\begin{equation}
H_{FL/DL}=\frac{\frac{\hbar}{2e}j_e}{\mu_0 M_S d_{FM}} \xi_{FL/DL}
\label{Eq:Hdlfl}
\end{equation}
where $\hbar$ is the reduced Planck's constant and $\xi_{FL/DL}$, the field-like and damping-like torque efficiencies obtained from the spin drift diffusion model \cite{Haney-2013, Chen-2013}, depending on the spin mixing conductance at the interface.




\section{Anisotropic and Spin-Hall magnetoresistances}
\label{sec:AMR}

\begin{figure}[htb]
\centering
\includegraphics[width=\textwidth, page=1, trim={0cm 16cm 0cm 0cm}, clip]{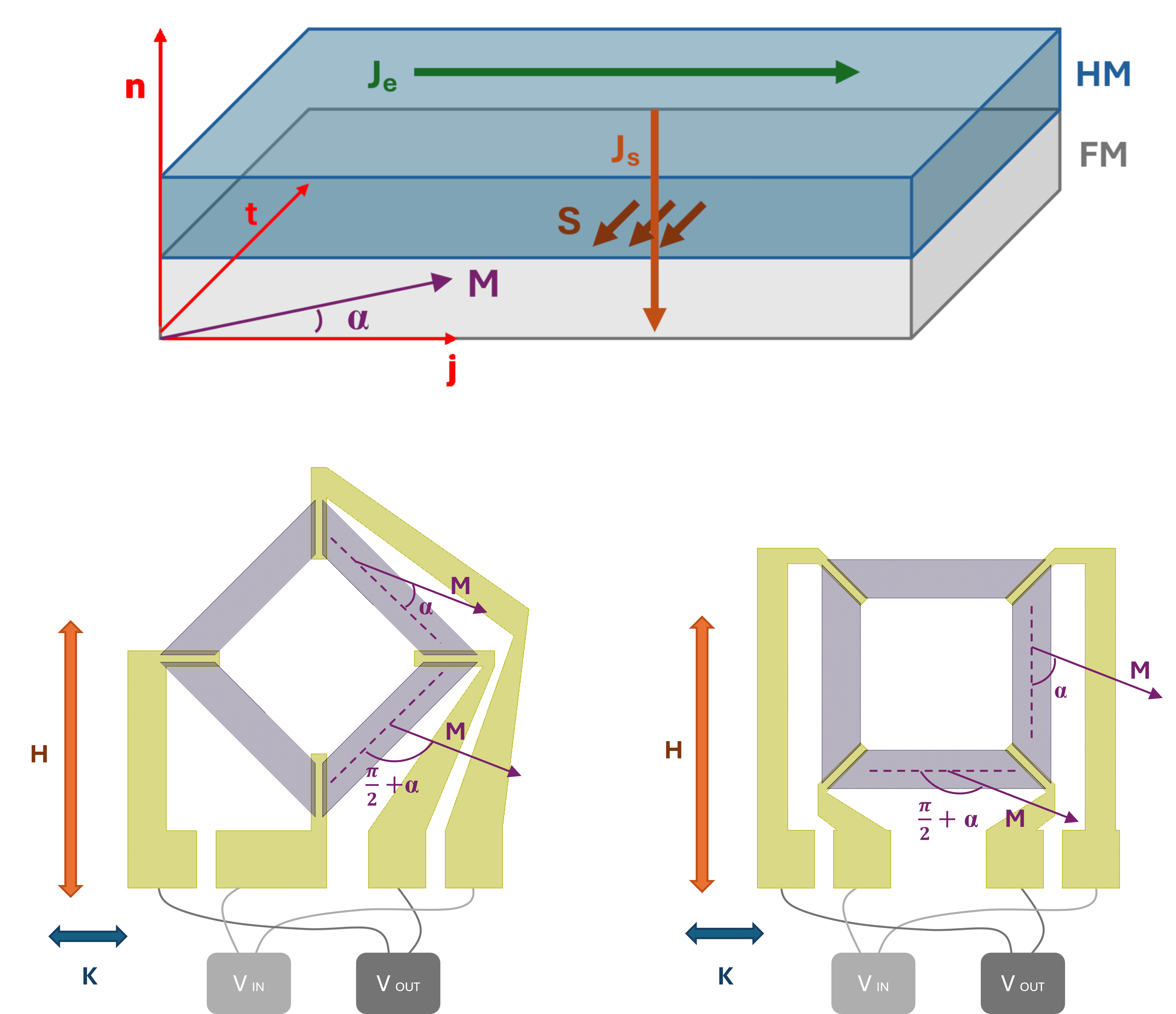}
 \caption{Spin current $J_\mathrm{s}$ injection in a bilayer system composed of a heavy metal (HM) and a ferromagnet (FM) due to spin Hall effect in presence of an electrical current $J_\mathrm{e}$.} \label{FIG:HMFM}
\end{figure}

Designing a device composed of conductive HM/FM  (see Fig.\ref{FIG:HMFM}) requires an understanding of the behavior of both AMR and SMR since part of the sensor current flows in the FM layer (where AMR is observed), while the other flows in the HM layer (where SMR can be observed).  \\

The AMR, observed for the first time in 1857 by lord Thomson \cite{McGuire-1975}, was later investigated by Smit \cite{Smit-1951, Marsocci-1965} and originates from the spin-orbit coupling (SOC) between the lattice and the spin system \cite{Silva_2015}, which allows for the mixing of majority and minority channels.
For a theoretical description, a two-current model approximation of conduction \cite{Mott-1936} by \textit{4s} electrons can be used below the Curie temperature, where the thermal spin wave density is low and its effect on spin mixing is negligible. 
The result is a scattering probability of electrons that depends on the angle between the direction of the electrical current and the magnetization with $R_\parallel > R_\perp$.
 As a consequence, AMR is maximum when the magnetization (whose direction is identified by the unit vector $\mathbf{m}$) is along the current direction $\mathbf{j}$, and minimum along the transverse ($\mathbf{t}$) and perpendicular to the plane ($\mathbf{n}$) directions. 

\begin{equation}
R_\mathrm{AMR}=R_\mathrm{FM} + \Delta R_\mathrm{AMR} m_j^2
\label{Eq:AMR}
\end{equation}
with $R_\mathrm{AMR}$ the resistivity minimum, $\Delta R_\mathrm{AMR}$ positive and $m^2_j=\cos^2 \alpha$, where $\alpha$ is the angle between the magnetization and the current.

On the contrary, the SMR originates from the reciprocal spin current $j_s$ transmission and reflection between the HM layer and the FM layer (not necessarily conductive) due to the spin-Hall effect \cite{Weiler-2012}. The absorption of $j_s$ depends on the orientation of $\mathbf{M}$, which gives rise to the total resistance dependence on the magnetization direction. As a result, the SMR is minimum when the magnetization is along the $\mathbf{t}$ direction and maximum along the $\mathbf{j}, \mathbf{n}$ direction. 

\begin{equation}
R_\mathrm{SMR}=R_\mathrm{HM} + \Delta R_\mathrm{SMR} m_t^2
\label{Eq:SMR}
\end{equation}
with $R_\mathrm{SMR}$ the resistivity maximum, $\Delta R_\mathrm{SMR}$ negative and $m^2_t=1-\cos^2 \alpha$. The SMR can be easily measured in a bilayer with $R_\mathrm{FM} \gg R_\mathrm{HM}$ by SOT-FMR \cite{Liu-2011}, harmonic Hall measurements \cite{avci2014} or angle-dependent MR \cite{Althammer-2013}. In fact, the first time it was observed in a bilayer where the FM layer was insulating YIG \cite{Weiler-2012}.
If this condition is not observed, we have to consider that the resistance of an HM/FM bilayer (Fig.\ref{FIG:WBridge}) is given by the parallel circuit of the two resistances $R_\mathrm{AMR}$ and $R_\mathrm{SMR}$, given by $R_\mathrm{AMR}=R_\mathrm{FM}+\Delta R_\mathrm{AMR}$ and $R_\mathrm{SMR}=R_\mathrm{HM}+\Delta R_\mathrm{SMR}$, resulting in:

\begin{align}
R & \approx R_{||}+\frac{R^2_\mathrm{HM}\Delta R_\mathrm{AMR}-R^2_\mathrm{FM}\Delta R_\mathrm{SMR}}{(R^2_\mathrm{HM}+R^2_\mathrm{FM})^2} \nonumber \\
R_{||}&=\frac{R_\mathrm{HM}R_\mathrm{FM}}{R_\mathrm{HM}+R_\mathrm{FM}}  
\label{Eq:Rtot}
\end{align}
With $\Delta R_\mathrm{AMR}$ and $\Delta R_\mathrm{SMR}$ given by Eqs.\ref{Eq:AMR},\ref{Eq:SMR}. Under the approximation of small $\Delta R_\mathrm{AMR}$, $\Delta R_\mathrm{SMR}$ we find that the parallel resistance is 

\begin{equation}
R \approx R_0+A \cos^2 \alpha
\label{Eq:Rcos2}
\end{equation}
where
\begin{align}
R_0&=R_{||}+\frac{R^2_\mathrm{FM} \Delta R_\mathrm{SMR}}{(R^2_\mathrm{HM}+R^2_\mathrm{FM})^2}  \nonumber \\
A&=\frac{R^2_\mathrm{HM}\Delta R_\mathrm{AMR}-R^2_\mathrm{FM}\Delta R_\mathrm{SMR}}{(R_\mathrm{HM}+R_\mathrm{FM})^2}
\label{Eq:Rcos2Params}
\end{align}

\begin{figure}[htb]
\centering
\includegraphics[width=\textwidth, page=1, trim={0cm 0cm 0cm 1.5cm}, clip]{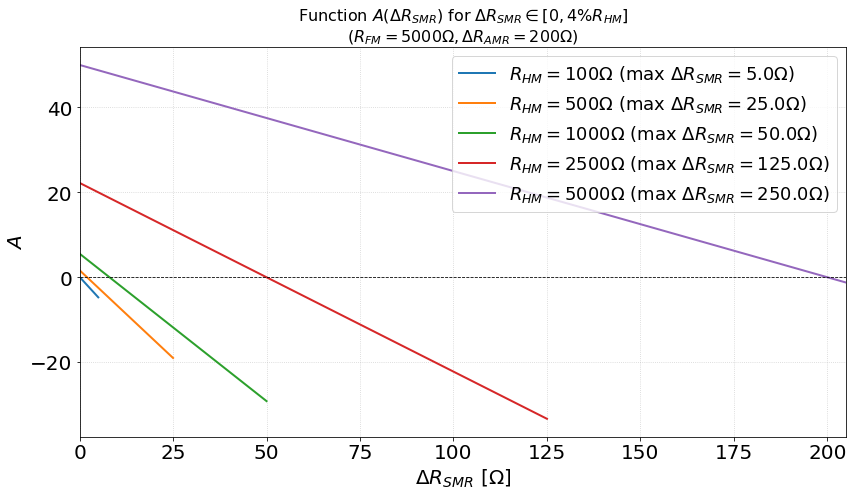}
\caption{The coefficient $A$ as a function of $\Delta R_{\mathrm{SMR}}$. The magnitude of $A$ depends on the ratio between $R_{\mathrm{HM}}$ and $R_{\mathrm{FM}}$. Here a fixed $R_{\mathrm{FM}}$=5000$\ohm$ is used with a fixed typical value $\Delta R_{\mathrm{AMR}}$ of 4\% of $R_{\mathrm{FM}}$. The values of $\Delta R_{\mathrm{HM}}$ are shown up to a maximum of 5\% of $R_{\mathrm{HM}}$ assuming the SMR to be of the same order as the AMR.}  
\label{FIG:A}
\end{figure}

The coefficient $A$ is an estimate of the maximum resistance change that can be obtained. It depends strongly on the relative resistances of the FM and HM layers. Fig.\ref{FIG:A} shows A as a function of $\Delta R_{\mathrm{SMR}}$ for different values of the HM resistance $R_{\mathrm{HM}}$. The $R_{\mathrm{FM}}$ is kept constant, and $\Delta R_{\mathrm{AMR}}$ is assumed to be 4\% of the value. Considering here the SMR effect being of the same order of magnitude (or less) as the AMR effect, we assume for $\Delta R_{\mathrm{SMR}}$ a maximum value of 5\% of the SMR, too. We observe that for $R_{\mathrm{HM}}$ comparable or larger to $R_{\mathrm{FM}}$ the $A$ coefficient is positive and the maximum value is obtained for $\Delta R_{\mathrm{SMR}}=0$. Therefore, highly sensitive sensor needs to be designed with $R_{\mathrm{FM}} > R_{\mathrm{HM}}$.

In the case that $R_{FM}\gg R_{HM}$, then from Eq.\ref{Eq:Rcos2Params} we can estimate $\Delta R_\mathrm{SMR}$ from the difference $A$ between the resistance saturation values for $\textbf{H} || \mathbf{j}$ and $\textbf{H} || \mathbf{t}$:

\begin{equation}
    A=R_{||}^2 \left( \frac{\Delta R_\mathrm{AMR}}{R^2_\mathrm{FM}} - \frac{\Delta R_\mathrm{SMR}}{R^2_\mathrm{HM}}\right) \approx - R_{||}^2 \frac{\Delta R_\mathrm{SMR}}{R^2_\mathrm{HM}} \approx - \Delta R_\mathrm{SMR}
    \label{Eq:RsmrSATS}
\end{equation}

The determination of the AMR and SMR contributions requires knowledge of the resistances of the single layers in the heterostructure. In order to determine these contributions accurately, the thickness dependent resistivity of the single layers needs to be considered. This can be done by employing a Fuchs-Sondheimer approach (see supplemental material). 

We note that the AMR and SMR contributions are not discernible in a measurement if $m$ stays in-plane, since $m_n^2=0 \Longrightarrow m_j^2=1-m_t^2$. If $m$ has a perpendicular-to-plane component, it is possible to discriminate between them: if $m$ rotates in the $(\mathbf{j},\mathbf{n})$ plane, there is no SMR angular dependence, while if $m$ rotates in the $(\mathbf{t},\mathbf{n})$ plane, there is an AMR angular dependence \cite{Althammer-2013}. Therefore, the only way to distinguish the MR contributions is an angle dependent measurement at saturated magnetization.

\FloatBarrier

\section{The Wheatstone bridge sensor model} \label{sect:WBmodel}
\subsection{Wheatstone bridge configuration}

\begin{figure}[htb]
\centering
\includegraphics[width=0.5\textwidth, page=1, trim={0cm 0cm 14cm 0cm}, clip]{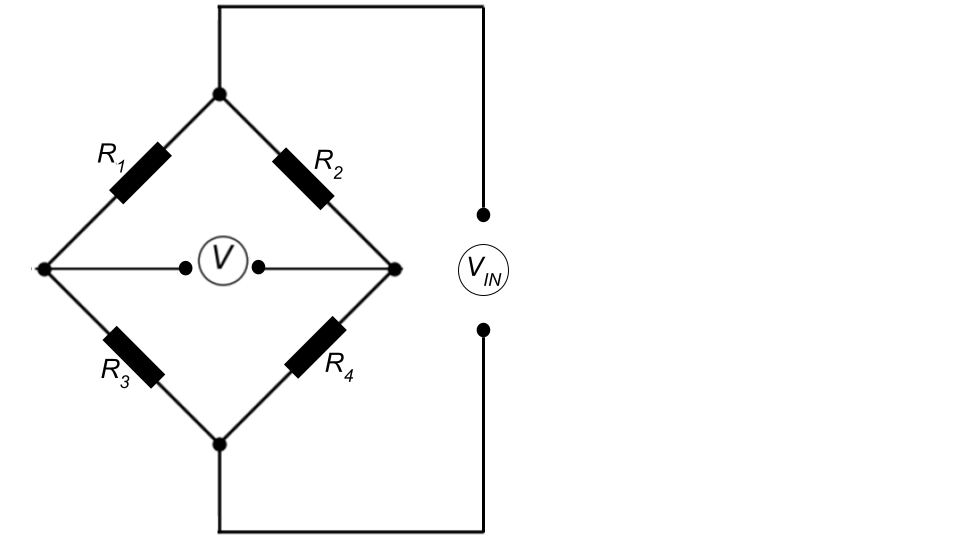}
 \caption{Sketch of a Wheatstone bridge resistance configuration.} \label{FIG:WBsketch}
\end{figure}

The Wheatstone bridge configuration (see Fig.\ref{FIG:WBsketch}) is the electrically equivalent to two parallel voltage dividers with a differential output. When balanced,  the output signal spans around zero, making it possible to use more precise measuring devices as it is easier to measure perturbations using a smaller measurement range. AMR-based sensors are typically implemented in this configuration to minimize common-mode noise and thermal drift, and to  convert very small resistance changes into to measurable differential voltage \cite{Demirci2020, dibbern1989}.

The Wheatstone bridge is designed in such a way that the output voltage ($V$) is  proportional to the input voltage ($V_\mathrm{in}$) and to relative  change  in total magnetoresistance ($\Delta R /R$) induced by  an external magnetic field $(V=V_\mathrm{IN}\frac{\Delta R}{R}$). The magnitude of the sensed voltage therefore depends on both the amplitude of the magnetoresistance  and the angle between the magnetization and current direction, as discussed in Eq.\ref{Eq:AMR}. In a typical AMR Wheatstone bridge sensor, the individual branches are barber poles, forcing the current to approximately 45$\degree$ with respect to the magnetization, thereby optimizing the balance between the magnetoresistance amplitude and reduced hysteresis. In the ideal case, the elements are single domain, so that the magnetization amplitude is constant. One of the advantages of the SMR sensor with respect to the AMR sensor is the fact that the resistance change is maximum when the current is perpendicular to the magnetization (see Eq.\ref{Eq:SMR}).


Here, we investigate and model a Wheatstone bridge configuration where the sensing principle is based on the mixed magnetoresistance effect, AMR and SMR, as proposed by \cite{Xu2018, Yang2017}. In the SMR bilayer composed of a HM and an FM thin film, the SOT can be exploited to substitute the barber pole structure and cause linearization by favoring a 45$\degree$ angle between magnetization and current, which is a compromise between the maximum AMR (along the current direction) and the least hysteresis (transverse to the easy anisotropy axis). As shown in Eq.\ref{Eq:SMR} the SMR signal is maximum in the transverse direction, which usually coincides with hard anisotropy axis (considering the current flowing along an elongated bridge element). 

Considering a simple Wheatstone bridge, where the resistances of the diagonally opposite branches (1,3) and (2,4) are equal ($R_1=R_3$ and $R_2=R_4$), the magnetization angle with $\mathbf{j}$ will be $\alpha$ along the 1,3 sides and $\pi/2+\alpha$ along the 2,4 sides, caused, for example, by SOT, as in Ref.\cite{Xu-SOT_2019}. 
We then obtain that the output voltage depends on the input voltage $V_\mathrm{IN}$ as $V=V_\mathrm{IN}\frac{R_1-R_2}{R_1+R_2}$. By making use of Eq.\ref{Eq:Rcos2} we have

\begin{equation}
V= \frac{I}{2}A \cos(2 \alpha)
\label{Eq:WBcos2}
\end{equation}
not dependent on $R_0$ and with $I$ the constant current fed into the bridge.

While in Ref.\cite{Xu-SOT_2019}, the FM film is Py, prototypical material for AMR, our aim is to model the most general case of an SMR wheatstone bridge type sensor, including magnetic anisotropy. Specifically, in CoFeB alloys, which are often employed for spintronic devices, magnetic anisotropy can be modified by heat treatment, composition, and texture, which depend on neighboring layers. In particular, our aim is to understand whether, under certain conditions, the magnetic anisotropy can enhance the effect of SOT and improve the linearization and performance of the sensor structure, forcing the magnetization in the transverse direction without causing strong hysteresis.

\subsection{Magnetization model - uniform magnetization}
The output voltage of the sensor structure as a function of the external magnetic field $V(H)$ depends on how the resistance changes with the magnetic field $R(H)$. In order to model this relation, a Stoner-Wohlfarth type model with uniaxial anisotropy can be applied in the first approximation, assuming that the single branches of the Wheatstone Bridge are single domain. This means that the magnetization amplitude remains constant, but its angle varies (see supplemental material). The anisotropy can be induced by the shape anisotropy of the elongated bridge branches or by a magnetic anisotropy of the FM layer defined during deposition process or thermal treatment.

To find the equilibrium magnetization angle with respect to the current direction, as a function of the external field, $\alpha=f(H)$, we need to write the energy equation $E/\mu_0 M_S=- \mathbf H_{\mathrm{eff}} \cdot \mathbf m$ with an effective field $\mathbf H_\mathrm{eff}$ including all magnetic terms according to the micromagnetic theory and $\mathbf{m}$ is the unit vector in the direction of the magnetization. The most common approach is to consider only the applied field $\mathbf H$ and the uniaxial anisotropy field $\mathbf H_K$:

\begin{equation}
    E/\mu_0 M_S=- \mathbf H \cdot \mathbf m - \mathbf H_K \cdot \mathbf m
    \label{EQ:en_SW}
\end{equation} 


Considering $\mathbf{H_\mathrm{K}}$ the magnetic anisotropy of the FM layer in an arbitrary direction (e.g. induced by field annealing) and an elongated Wheatstone bridge branch with size (L,w,$d_{\mathrm{FM}}$), we can add a demagnetizing field along the long axis $\mathbf{L}$ ($w$ is the short axis and $d_{\mathrm{FM}}$ the film thickness) approximated by \cite{Hubert-1998}:

\begin{equation}
    \mathbf{H_\mathrm{d}} = - M_S \frac 1 2 \mathbf{L} w t_{FM} \int_0^\infty [(L^2+\eta)\sqrt{(L^2+\eta)(w^2+\eta)(t^2_{FM}+\eta)}]^{-1} d\eta
\end{equation}


The current flow generates additional fields that influence the magnetization direction, the magnetic field generated by the current $\mathbf H_{\mathrm{Oe}}$ and the effective torque fields $\mathbf H_{\mathrm{DL}}$ and $\mathbf H_{\mathrm{FL}}$.

The Oersted field $\mathbf H_{\mathrm{Oe}}$ can be calculated in the general case from the Karlqvist formula \cite{karlqvist-1954}. Here, in the first approximation, the field generated within the DC current carrying layer is sufficient, given by $H_{Oe}=d_{\mathrm{HM}} j_{\mathrm{HM}}/2$ and $H_{Oe}=d_{\mathrm{FM}} j_{\mathrm{FM}}/2$ where $d_{\mathrm{HM/FM}}$ are the thicknesses of the two layers and $j_{\mathrm{HM/FM}}$ are the current densities in the two layers.

While the Oersted field adds to the effective field in Eq.\ref{EQ:en_SW}, the field-like and damping-like torque fields do not directly contribute to the energy equation, since both are always perpendicular to the magnetization $\mathbf M$: $\mathbf H_{FL}=H_{FL} \mathbf m \times \mathbf s$, $\mathbf  H_{DL}=H_{DL} \mathbf m \times (\mathbf m \times \mathbf s)$. They derive from external forces (the spin current) and must be taken into account as torques in the LLG equation (see Eq.\ref{Eq:LLGstat}) with an effective field $\mathbf H_{eff}=\mathbf H+\mathbf H_K+ \mathbf H_d+\mathbf H_{Oe}$.

We thus obtain, imposing $d \mathbf{M}/dt=0$:

\begin{equation}
    \mathbf{M} \times \mathbf{H}_{\mathrm{eff}} = \frac 1 {\mu_0 \gamma_G} \frac{\mu_B M_S j_{HM}}{e d_{FM}}(\xi_{FL} \mathbf{m}\times \mathbf{s} +\xi_{DL} \mathbf{m}\times(\mathbf{m}\times \mathbf{s}))
    \label{Eq:LLGstat}
\end{equation}
where $\mathbf{s}=-\mathbf{t}$ is the spin current polarization direction.

The resulting system of equations (see supplemental material) can be numerically solved for the magnetization angle in-plane $\alpha$ and perpendicular-to-plane $\theta$, from which the Wheatstone bridge voltage output can be calculated.


\subsection{Magnetization model - nonuniform magnetization}
\label{sect:numodel}

In the case of real sensor devices, the assumption of a uniform magnetization configuration is generally true, since the magnetic properties and device geometries are chosen to fulfill this criterion. However, for studying and designing SOT devices in general, and sensors in particular, it is interesting to include the effect of magnetization reversal through domain wall motion. In micrometer-sized Hall bars or bridge branches, magnetic domains might be present depending on the magnetization and anisotropy of the FM material. We, therefore, employ a more complete approach considering both domain wall (DW) motion and coherent rotation, based on the Stoner-Wohlfarth model \cite{Bertotti-1998}. In this approach we assume an ensemble of non-interacting elementary single domain units ("particles"), saturated (with local magnetization $\mathbf M=M_S \mathbf{m}$) and characterized by a local anisotropy axis with constant $K_i$ \cite{Bertotti-1998}. The antagonism between anisotropy and Zeeman energy drives the reversible coherent rotations of $\mathbf M$ and triggers its irreversible jumps (Barkhausen jump \cite{barkhausen1919}). In the Stoner-Wohlfarth framework it is possible to include DW assisted magnetization reversal, occurring in correspondence of a local switching field lower than the local anisotropy field \cite{Appino-2023, Appino-2000}. In the concrete case we assume a constant value $K_i$, with the local anisotropy axes with angles $\varphi_i$ homogeneously distributed, in addition to a constant uniaxial anisotropy $K_u$. It is possible to account for this additional energy contribution by associating with each region, originally characterized by the couple of values ($K_i$,$\varphi_i$), an effective easy axis with an anisotropy constant $K=K(K_i,\varphi_i;K_u)$ forming an angle $\varphi = \varphi(K_i,\varphi_i;K_u)$ with $K_u$, according to the relationships

\begin{align}
K &= \sqrt{ K_i^2 + K_u^2 + 2K_iK_u\cos 2\varphi_i } \nonumber \\
\tan 2\varphi &= \frac{ \sin 2\varphi_i}{ \cos 2\varphi_i + K_u/K_i} 
\label{eq:KAPPA}
\end{align}

\begin{figure}[htb]
\centering
\includegraphics[width=0.5\textwidth]{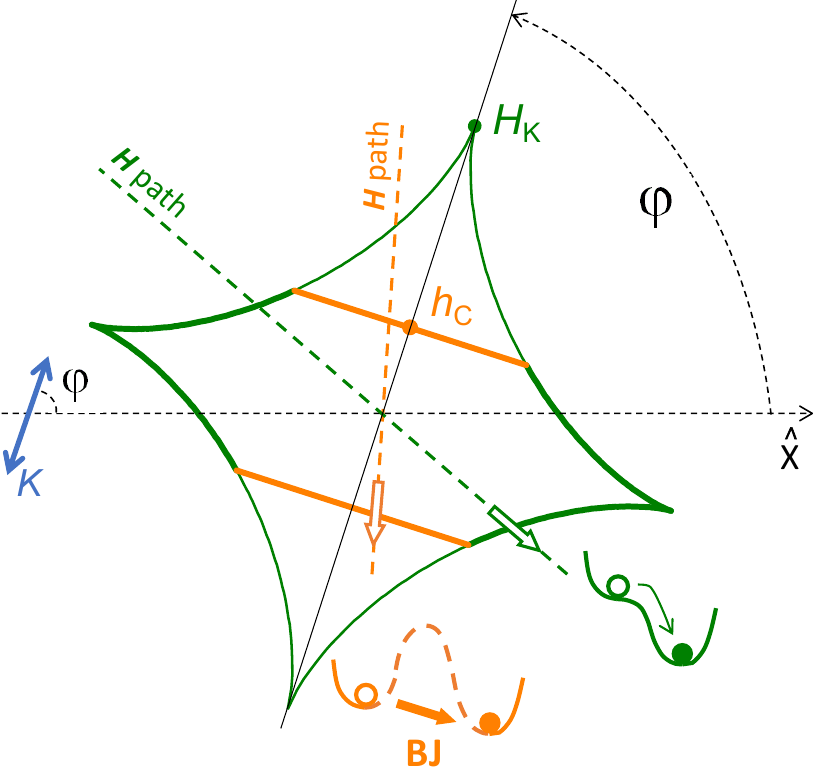}
\caption{Truncated astroid (orange threshold) representing the modified Stoner-Wohlfarth model of a single "particle" in the ensemble of statisically distributed ones, characterized by an easy axis with anisotropy constant $K$ forming an angle $\varphi$ with $\mathbf{x}$. Two $\textbf{\textit{H}}$ paths not involving (olive) or involving (orange) the threshold corresponding to a local Barkhausen jump are shown. Note the difference between the two Gibbs free energy double-well profiles corresponding to the point where $\textbf{\textit{H}}$ crosses the classical (thick olive) or truncated (thick orange) threshold.}  
\label{FIG:model}
\end{figure}

As proposed in \cite{Appino-2000}, the role played by the DW displacement in the magnetization reversal can be described by means of a "\textit{truncated} Stoner-Wohlfarth astroid" that describes the behavior of each of these ($K,\varphi$) particles. As illustrated in Fig.\ref{FIG:model}, the magnetization process is qualitatively described following the line $H_\mathrm{path}$, so the evolution of the magnetization with an applied field. For each particle $H_\mathrm{path}$ can cross either the "classical", i.e., not truncated, astroid border (olive), exiting it, or a new threshold (orange) characterized by a local coercive field $h_c = T H_{K}$, with $H_{K}=2K/\mu_0 M_s$ being the effective local anisotropy field, and the constant $0< T \le 1$ quantifying the relative importance of the DW processes. In the first case, irreversible rotation occurs, whereas the second one gives rise to a Barkhausen jump: an irreversible displacement of a small DW portion. In correspondence of the classical or truncated astroid crossing, the Gibbs free energy double-well profile is also shown. In the second case, one can notice that the presence of a DW allows the system to occupy the lowest energy level when $\textbf{H}$ exceeds a threshold identified by a line perpendicular to the particle local anisotropy axis at a threshold $h_c$.

Note that for $T$=1, only (reversible or irreversible) rotations are responsible for the magnetization process. It must also be remarked that the coherent rotation behavior does not change if the classical or truncated astroid governs the process.

With this model, it is possible to obtain numerically the $M(H)$ curves of the SOT device in the presence of domain wall processes, with a statistical distribution of single domain particles. As a consequence, we can derive the magnetoresistance as a function of the magnetic field.

\FloatBarrier

\section{Validation of the model}\label{sect.Experimental}

\subsection{The Wheatstone bridge test device}

\begin{figure}[htb]
\centering
\includegraphics[width=\textwidth, page=1, trim={0cm 0cm 0cm 12cm}, clip]{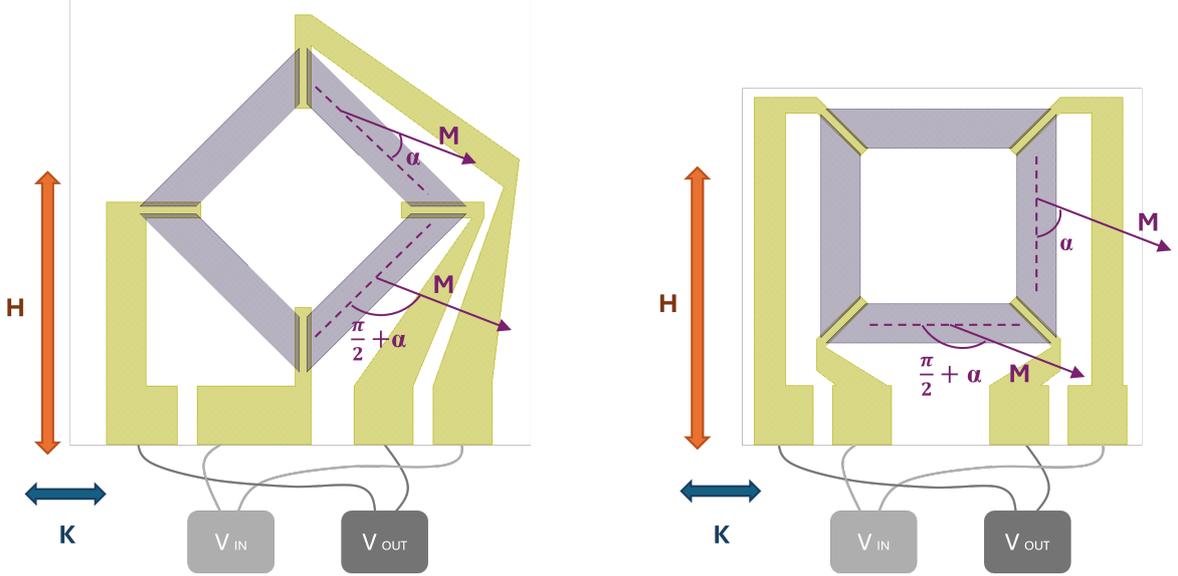}
 \caption{The two sensor configurations which were modeled, pattern \textbf{p45} at 45\degree (left) and pattern \textbf{p0} parallel (right) with respect to the applied magnetic field $H$. $K$ indicates the direction of an eventually present magnetic anisotropy.} \label{FIG:WBridge}
\end{figure}

To test the model and validate its proper functioning, it was adapted to a realistic Wheatstone bridge configuration that is simple to fabricate and measure \cite{renaudin2010}. The test device consists of a squared Wheatstone bridge of different dimensions. Two different configurations with respect to an assumed magnetic anisotropy were studied: \textbf{p45} with long axis of a rectangular branches at 45$\degree$ with respect to the magnetic anisotropy and \textbf{p0} with long axis parallel to the anisotropy (see Fig.\ref{FIG:WBridge}).

For the uniform magnetization model, for which we can derive an analytical solution, we distinguish between the two cases for the two sensor configurations \textbf{p45} and \textbf{p0} and obtain the energy equations from Eq.\ref{EQ:en_SW}:

\begin{align} 
E_\mathbf{p45}/\mu_0 M_S&=\frac {\sqrt{2}} 2 H(\cos \alpha - \sin \alpha)-\frac{H_K}{2}(\cos \alpha+\sin \alpha)^2  \nonumber \\ 
E_\mathbf{p0}/\mu_0 M_S&=- H \sin \alpha-H_K \cos^2 \alpha
\label{Eq:EnModel0}
\end{align}
with the stability condition is 
\begin{equation}
    \alpha=\begin{cases}
        \alpha_0+arcsin(H/2H_K) & \textbf{if } |H|<2H_K \\
        \alpha_0+\pi/2 \cdot sign(H)  & \textbf{if } |H|>2H_K 
    \end{cases}
\label{Eq:alphaVsH_SW}
\end{equation}
with $\alpha_0=0$ for \textbf{p0} and $\alpha_0=\pi/4$ for \textbf{p45}. The $\pi/4$ displacement indicates a linear response of the \textbf{p45} sensor around $H=0$, as desired for sensor devices.


We then obtain the Wheatstone bridge voltage in the two configurations from Eq.\ref{Eq:WBcos2}:

\begin{equation}
V_\mathbf{p45}/I=\begin{cases}
    \frac A 2 \frac H {H_K} \sqrt{1- \frac{H^2}{4H^2_K}} & \text{if} |H|<2H_K\\
    0 & \text{if} |H|>2H_K
\end{cases}
\label{Eq:Vout45}
\end{equation}

\begin{equation}
V_\mathbf{p0}/I=\begin{cases}
    \frac A 2 \left(1- \frac{H^2}{2H^2_K}\right) & \text{if} |H|<2H_K\\
    -A/2 & \text{if} |H|>2H_K
\end{cases}
\label{Eq:Vout0}
\end{equation}

\subsection{Sample fabrication}

The simple Wheatstone bridge sensors were fabricated from 100~mm wafers of HM(thickness $d_{HM}$)/Fe$_\mathrm{60}$Co$_\mathrm{20}$B$_\mathrm{20}$(2~nm) bilayers with a 2~nm insulating capping Ta layer (oxidized in the ambient atmosphere). FeCoB was chosen because it is a typical material choice for spintronic sensors, and the 2~nm thickness guaranties sufficiently high resistance and an in-plane magnetic anisotropy, which is desired in conventional  MR sensors. The bilayers were sputter deposited at Singulus Technologies AG by a Timaris cluster tool system on oxidized Si wafers. For the Wheatstone bridge devices, HM=Pt and Ta (5~nm) were employed, while Hall bars with variable HM thicknesses in the range of 2-15 nm were used. In order to study the influence of magnetic anisotropy on sensor performance, the samples were annealed in a magnetic field of 1 T for 2 hours at 300 \degree C. The saturation magnetization was obtained from ferromagnetic resonance measurements of the films and is $\mu_0 M_S \approx 1.2~ T$. \cite{skowronski-2021}.

\begin{figure}[htb]
\centering
\includegraphics[width=12cm]{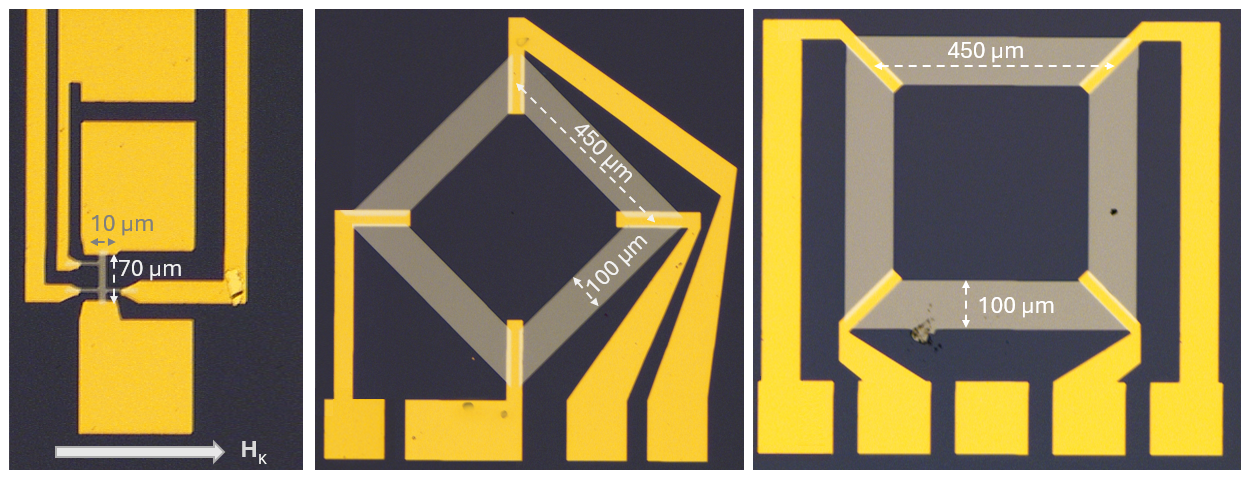}
 \caption{Left: Hall bar pattern; center and right: Wheatstone bridge design. The images where taken with an optical microscope.} \label{FIG:patterns}
\end{figure}

The multilayers were patterned using optical lithography and ion-beam etching, in the shape of Hall bars with the anisotropy axis perpendicular to the long axis of the Hall bar, and of Wheatstone bridge devices both parallel and at $45 \degree$ to the anisotropy axis (Figure \ref{FIG:patterns}). The Hall bars are $L=70~ \mu m$ long and $w=10~ \mu m$ wide; the Wheatstone bridge branches are $L=450~ \mu m$ long and $w=100~ \mu m$ wide.

Finally, the second lithography step, followed by a Ti(5)/Au(50) deposition and lift-off process, was used to pattern contact pads design to be compatible with our rotating probe stations in order to characterize the fabricated devices. 

\subsection{Magnetoresistance measurements}

\begin{figure}[htb]
\centering
\includegraphics[width=\textwidth]{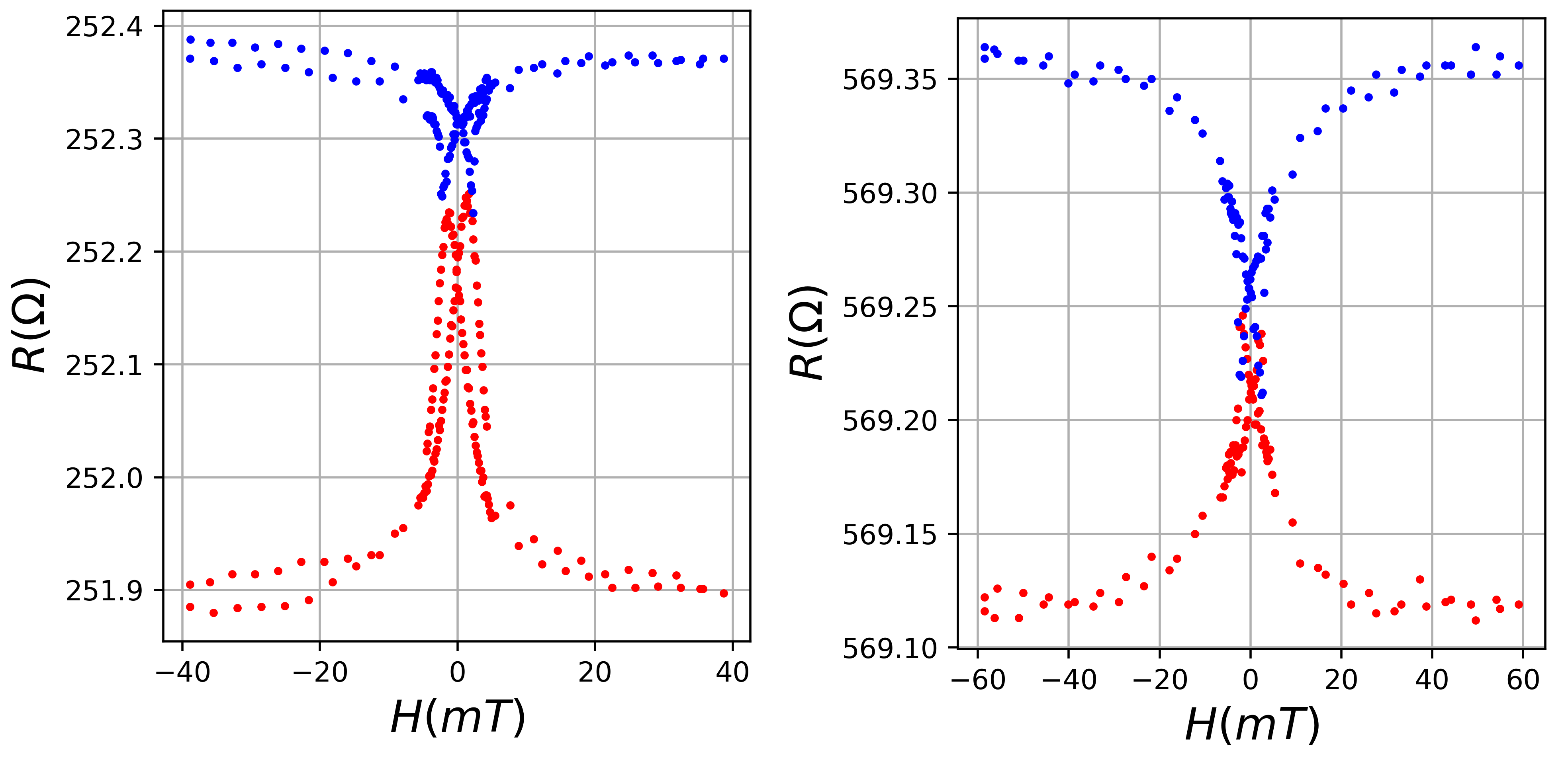}
\caption{Longitudinal magnetoresistance measurements of HM(5~nm)/Fe$_\mathrm{60}$Co$_\mathrm{20}$B$_\mathrm{20}$(2~nm) patterned into Hall bars with the field parallel (blue) and transverse (red) to the Hall bar. Left: HM=Pt; Right: HM=Ta} \label{FIG:RvsH}
\end{figure}

Longitudinal magnetoresistance measurements (Fig. \ref{FIG:RvsH}) were performed in an electromagnet on the  bilayers (HM(5~nm)/Fe$_\mathrm{60}$Co$_\mathrm{20}$B$_\mathrm{20}$(2~nm)) patterned into Hall bars shown in Fig. \ref{FIG:patterns}, using a Keithley 2636 A source-meter and Stanford SR830 DSP Lock-in Amplifier. Typical magnetoresistance curves for the field applied parallel ($R_j$, blue curve) and transverse ($R_t$, red curve) to the electrical current are presented with a constant current of $100\,\mu\text{A}$ (for HM=Pt) and $1\,\text{mA}$ (for HM=Ta). The resistance difference $R_{j,\text{max}} - R_{t,\text{min}}$ corresponds to the factor $A$ in Eq. \ref{Eq:RsmrSATS}, which is a combination of SMR and AMR contributions. Similar measurements on Hall bars with different HM thicknesses were used to characterize the torque fields arising in the structures \cite{Magni2022}.

It is often argued that these contributions can be distinguished by measuring with an applied field parallel and perpendicular to the current. However, we note that this is an oversimplified picture \cite{Choi-2017}; it is generally not possible to isolate the SMR and AMR magnitudes or relate the peak height directly to their individual amplitudes through these measurements alone (and in fact angle dependent measurements out of plane are required). This is especially true during magnetization reversal near the coercive field, where the observed resistance peaks are influenced by complex domain configurations. In this regime, the system cannot be interpreted simply as a parallel combination of two static magnetoresistance contributions. In fact, the peak height reflects both the intrinsic SMR/AMR contributions and the specific domain processes governing the magnetization reversal. If the process consisted of pure coherent rotation, measuring along the easy axis, the magnetoresistance would remain constant in the high resistance state, as the magnetization would switch instantaneously at the coercive field without changing its relative angle to the current. The emergence of peaks can instead be interpreted as the appearance of reversal domains with magnetization components along the hard axis (perpendicular to the current), which correspond to a lower resistance state. Consequently, both the coercive field and the volume fraction of these perpendicular domains determine the final shape and height of the magnetoresistance peaks, together with the SMR and AMR magnitudes. To get deeper insight into the microscopic process of magnetization reversal, magneto-optical measurements were conducted.

\FloatBarrier
\subsection{Domain configuration and magnetization modeling} \label{sect:MOKE}
 
\begin{figure}[htb]
\centering
\includegraphics[width=\textwidth]{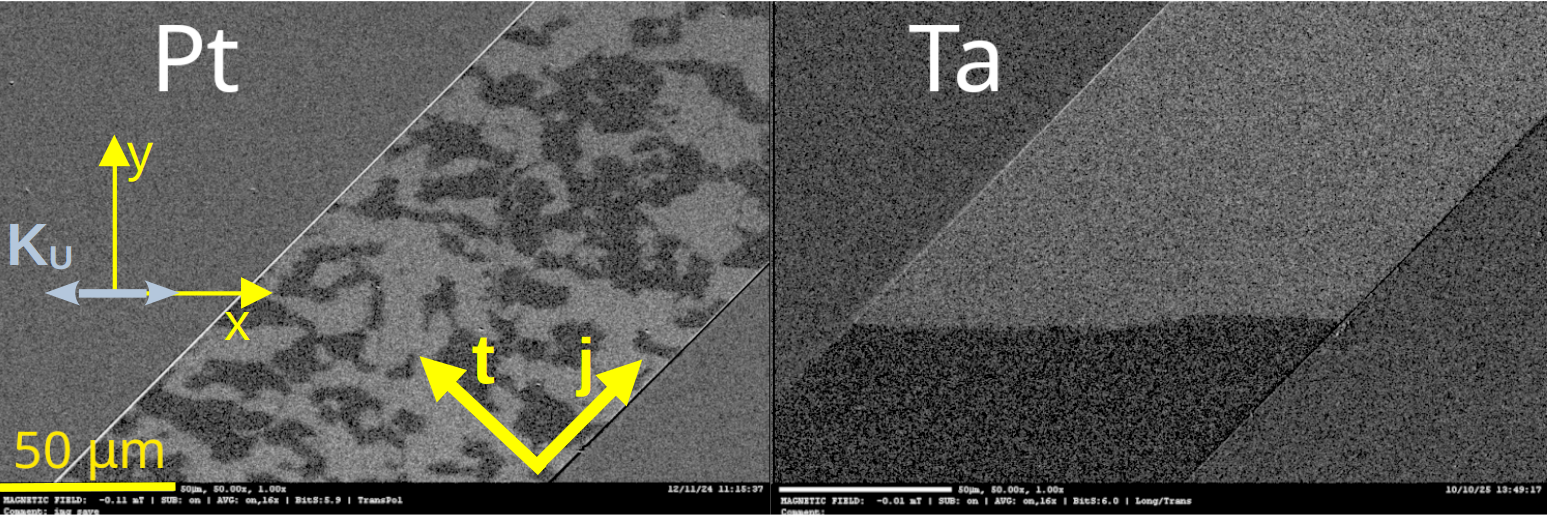}
\caption{Domain structure in the demagnetized state for the Pt and Ta sample. In the left image both the ($\mathbf{x}$-$\mathbf{y}$) and the ($\mathbf{j}$-$\mathbf{t}$) reference frames and the uniaxial magnetic anisotropy direction ($K_u$) are indicated.}
 \label{FIG:Domains}
\end{figure}

\begin{figure}[htb]
\centering
\includegraphics[width=\textwidth]{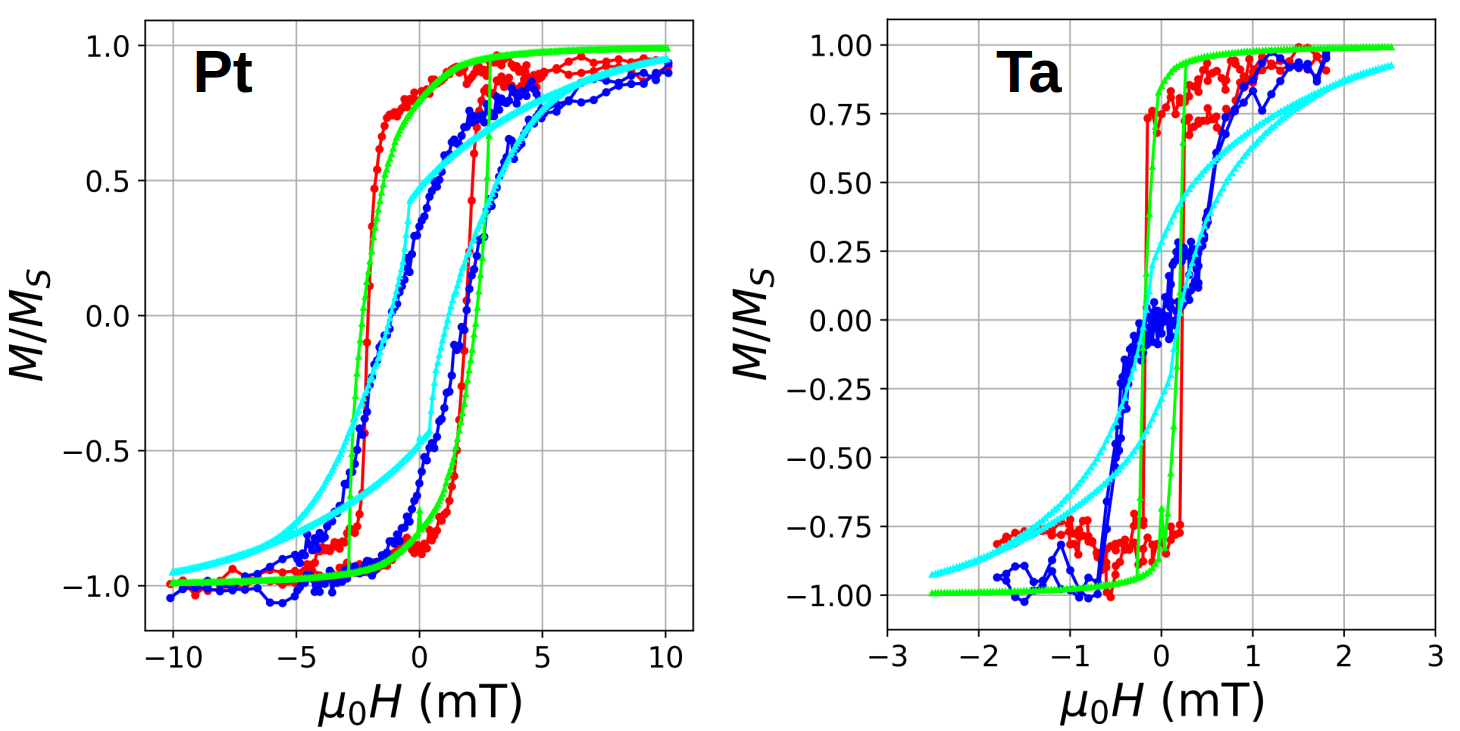}
\caption{Hysteresis loops the for Pt (left) and Ta (right) sample. The dots represent the measured magneto-optical loops. In red the measure along the E.A., with the applied field and the Kerr sensitivity both along $\mathbf{x}$, while in blue is shown the measure along the H.A. along $\mathbf{y}$. The lines correspond to the fits (respectively green and cyan) given by the modified Stoner-Wohlfarth model exploiting the "truncated" astroid.}  
\label{FIG:MOKEloops}
\end{figure}

Magneto-optic Kerr (MOKE) imaging was performed using a high-resolution microscope (Evico Magnetics) equipped with 20x, 50x and 100x (oil immersion) objective lenses. The imaging was employed to observe the domain configuration during magnetization reversal and to obtain qualitative magnetization curves $M(H)$.
For the acquisition of the magnetization curves of the HM(5~nm)/Fe$_\mathrm{60}$Co$_\mathrm{20}$B$_\mathrm{20}$(2~nm) samples, the magnetizing field is applied either along the $\mathbf{x}$ or the $\mathbf{y}$ direction (red axes, Fig. \ref{FIG:Domains}), obtaining the $M_x(H_x)$ or the $M_y(H_y)$ loops (red and blue dots in Fig.\ref{FIG:MOKEloops}), respectively. The $\mathbf{x}$ direction is aligned with the direction of the field-annealing, which induces a macroscopic easy axis with an anisotropy constant $K_u$.
The samples were later demagnetized with the field oscillating from $\pm$30 mT at a frequency of 23 Hz and decaying with a time constant of 3 seconds.
The Kerr domain imaging of the $M_x$ component is shown in Fig. \ref{FIG:Domains} for the two samples with different HM materials after demagnetization. 
Clearly, neither of the samples follow the simplified assumption of uniform rotation of the magnetization. The Pt-based sample shows a complex domain configuration, wheras in the Ta-based device, magnetization  reversal occurs through the motion of a single or low number of domain wall. An approximate estimate of the total domain wall density  is defined as the total domain wall length divided  by  the observed sample area. This density  was obtained by image processing and results for the Pt-based sample  $0.17 \pm 0.01 ~\mu m^{-1}$ and $0.01 \pm 0.003 \mu m^{-1}$ for Ta-based device. The corresponding total domain wall length is $3000 \pm 200 \mu m$ (Pt) and $200 \pm 60 \mu m$ (Ta) over an area of $17500 \mu m^2$.

To further elucidate the micromagnetic process, a MOKE hysteresis loop for two HM-samples was performed. The behavior of the experimental loops, shown in Fig.\ref{FIG:MOKEloops} by red and blue dots, is interpreted in the framework of the modified Stoner-Wohlfarth model (section \ref{sect:numodel}). The $M_x(H_x)$ and $M_y(H_y)$ loops, calculated by integration over the whole particle ensemble as described in the modeling section, are included in Fig. \ref{FIG:MOKEloops} for the Pt and Ta samples as cyan and orange lines.
From the $M_y(H_y)$ slope, it is possible to estimate the uniaxial anisotropy $K_u$ value as 1600 J/m$^3$ (Pt) and 600 J/m$^3$ (Ta). The remaining parameters are the local anisotropy $K_\mathrm{i} = \mathrm{2200~J/m}^3$ (Pt), $\mathrm{500~J/m}^3$ (Ta) and the parameter $T=0.3$ (Pt), $0.1$ (Ta). The difference in the $T$ values is related to different domain size present during the magnetization reversal, as depicted in Fig.\ref{FIG:Domains}.
The average anisotropy given by the model can be calculated as $\langle K\rangle=\frac 1 \pi \int_{-\pi/2}^{\pi/2} K(\varphi_i) d \varphi_i$, with $K(\varphi_i)$ given by Eq.\ref{eq:KAPPA}, resulting in $\langle K\rangle=2500~\mathrm{J/m^3}$, $H_K=16.5~\mathrm{mT}$ (Pt) and $\langle K\rangle=710 ~\mathrm{J/m^3}$, $H_K=4.7 ~\mathrm{mT}$ (Ta). We find that the Ta sample is significantly softer than the Pt sample and the magnetization reversal is mainly through domain wall motion. This difference between the two samples is attributed to the stronger interface intermixing in the Ta-sample, as was observed in the past \cite{cecot_influence_2017} comparing to the Pt-sample \cite{skowronski-2021}. The feature observed in the experimental hysteresis curve around the remanence in the $\mathbf{y}$ direction may be related to the nucleation of the domain wall and is not reproduced by the modified Stoner-Wohlfarth model. Nevertheless, an overall good agreement is reached.

\FloatBarrier

\subsection{Wheatstone bridge Characterization}

\begin{figure}[htb]
\centering
\includegraphics[width=10cm]{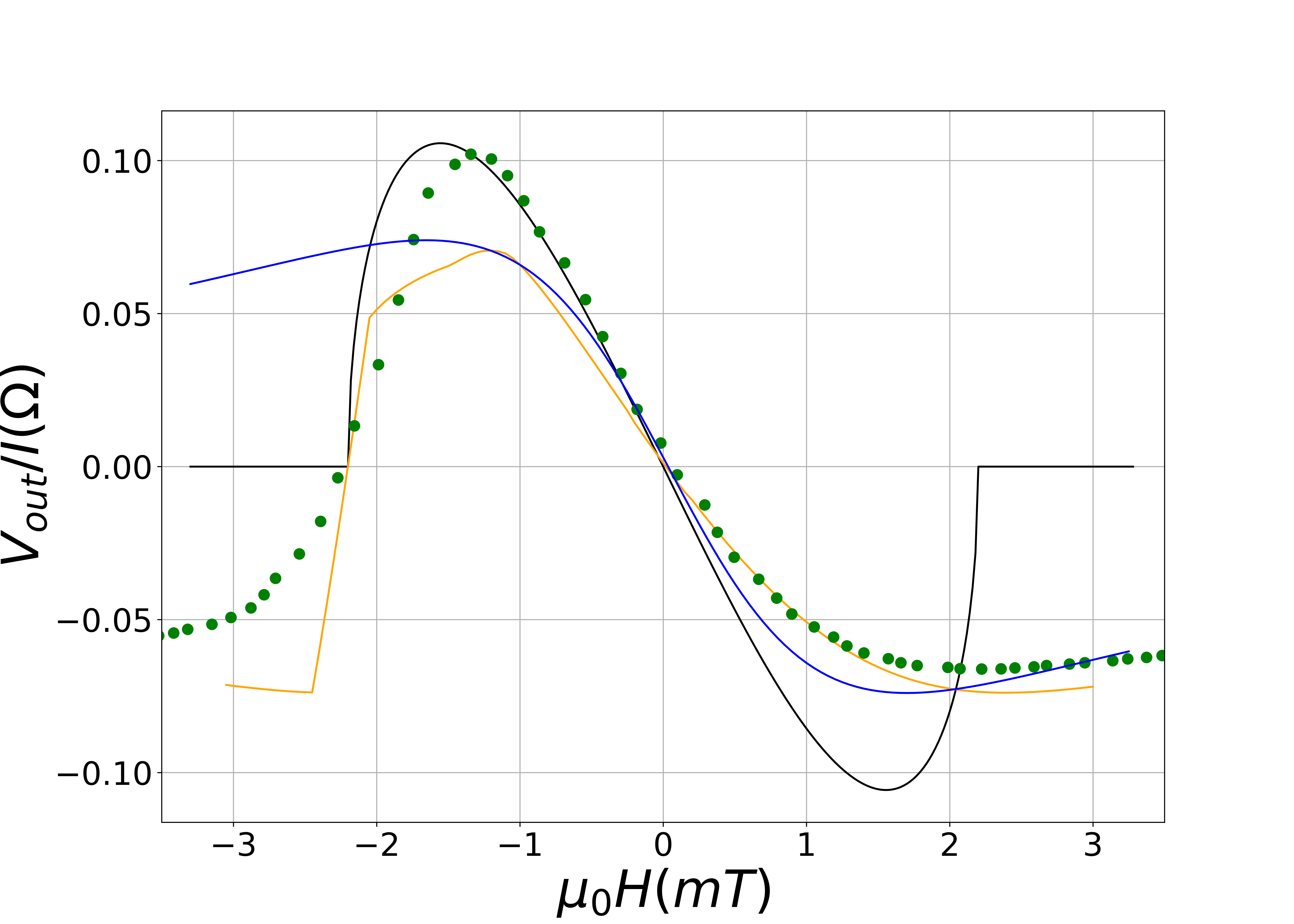}
\caption{
Pt sensor in \textbf{p45} configuration, Wheatstone bridge output as a function of the applied field, with constant current $I=5 mA$ (green dots). Predictions from the Stoner-Wohlfarth model without (black line) and with misalignment of an angle $\beta=0.3$ (orange line) use $\mu_0 H_K=1.1~\mathrm{mT}$; prediction with the LLG model from Eq.\ref{Eq:WBcos2} with the magnetization position obtained by numerical integration  (blue line) uses $\mu_0 H_K=1.7 mT$. Parameters used are  $\Delta R_{\mathrm{SMR}}= -250 \mathrm{m}\ohm$,  $\Delta R_{\mathrm{AMR}}= 190 \mathrm{m}\ohm$ and $R_{||}$= 312.8 $\ohm$.
}  \label{FIG:WB_Pt}
\end{figure}

\begin{figure}[htb]
\centering
\includegraphics[width=10cm]{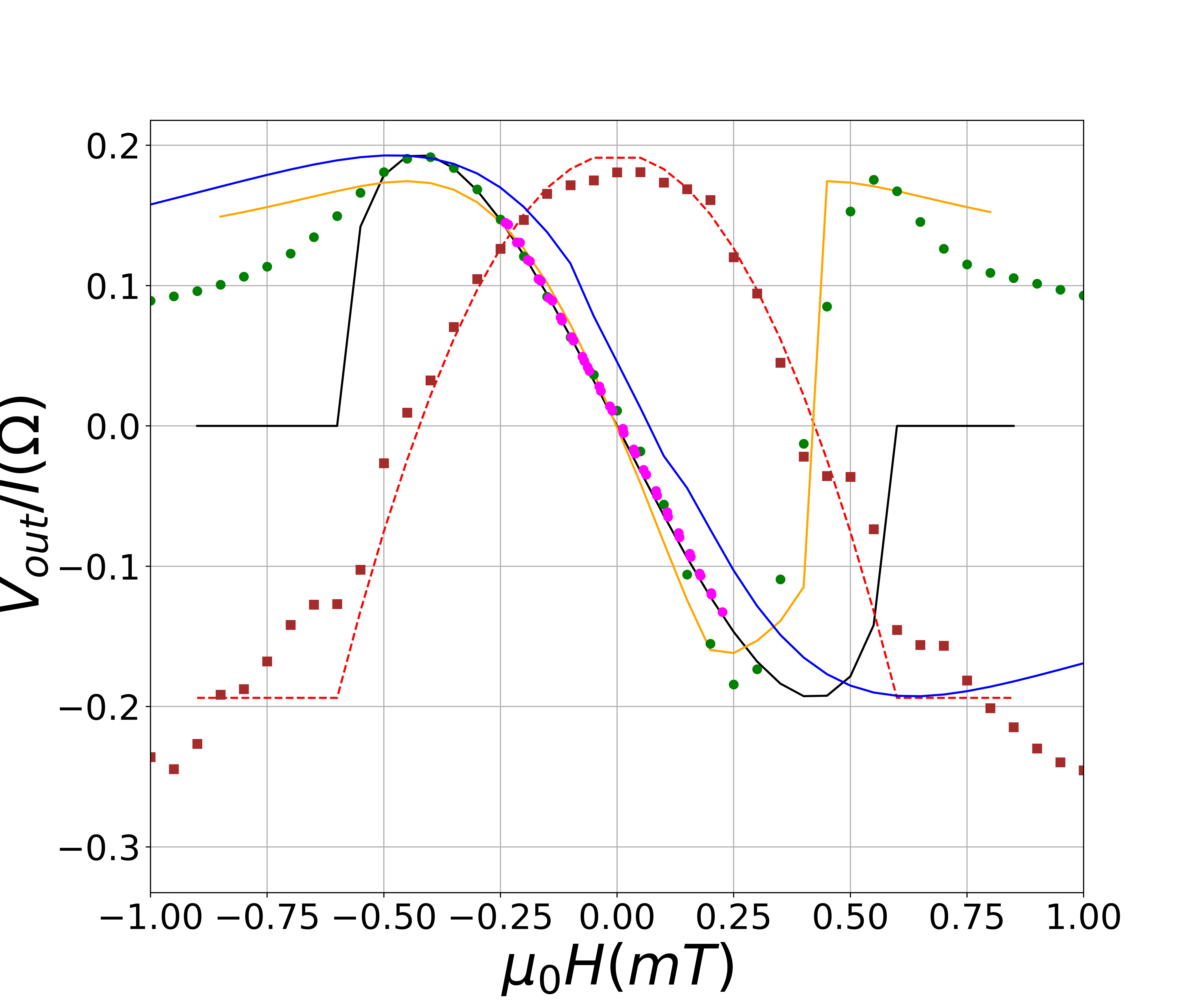}
\caption{Ta sensor: Wheatstone bridge output as a function of the applied field, with constant current $I=5 ~mA$. The data acquired in the \textbf{p45} configuration are green dots, magenta dots show the data points in a lower field range. The data acquired with the sensor in the \textbf{p0} configuration are the brown dots.
The uniform Stoner-Wohlfarth model for the \textbf{p45} configuration (black line) and the \textbf{p0} configuration (red dashed line). The orange curve shows the \textbf{p45} configuration including a misalignment of $\beta=0.3$. The LLG model is shown as the blue line. Parameters used are  $\Delta R_{\mathrm{SMR}}= -730 \mathrm{m}\ohm$,  $\Delta R_{\mathrm{AMR}}= 670 \mathrm{m}\ohm$ and $R_{||}$= 1331.5 $\ohm$.
}  \label{FIG:WB_Ta}
\end{figure}

Finally, the Wheatstone bridge sensors were characterized using a constant current $I = 5\text{ mA}$ and compared with  the numerical models prediction. Figures \ref{FIG:WB_Pt} and \ref{FIG:WB_Ta} show  the measured output voltage versus the applied magnetic field alongside simulation results . In Fig. \ref{FIG:WB_Pt}, the green dots represent the experimental data for the Pt-based sensor in the \textbf{p45} configuration (branches at $45\degree$ to the induced anisotropy), while the solid lines depict the simulation results.
The black line corresponds to the analytical Stoner-Wohlfarth solution (from Eq.\ref{Eq:Vout45}), utilizing the anisotropy field $H_K$ as the sole free parameter. We attribute the observed discrepancy between the basic model and experimental data to a slight misalignment of the uniaxial anisotropy axis relative to the applied field, likely occurring during either the magnetic field-annealing or device mounting for measurement. To account for this, the energy equations \ref{Eq:EnModel0} were modified to include a misalignment angle $\beta$ between $\mathbf{H}$ and $\mathbf{H_K}$:

\begin{align} 
E_\mathbf{p45}/\mu_0M_S&=\frac {\sqrt{2}} 2 H(\cos(\alpha-\beta)- \sin(\alpha-\beta))-\frac{H_K}{2}(\cos \alpha+\sin \alpha)^2  \nonumber \\ 
E_\mathbf{p0}/\mu_0M_S&=- H \sin(\alpha-\beta)-H_K \cos^2 \alpha
\label{Eq:EnModelbeta}
\end{align}

As shown by the orange line in Fig. \ref{FIG:WB_Pt}, a misalignment of $\beta = 0.3^\circ$ provides excellent agreement with the experimental behavior. Additionally, the LLG model (from Eq.\ref{Eq:WBcos2}) was employed to determine the magnetization position through numerical integration (blue line), which shows a reasonable agreement, particularly in the linear regime.

The robustness of this approach was further verified by modeling both \textbf{p45} and \textbf{p0} configurations for the Ta-based sensor (described by Eqs.\ref{Eq:Vout45} and \ref{Eq:Vout0}), with results shown in Fig. \ref{FIG:WB_Ta}. Very good agreement has been reached between the experimental data and the model for the two cases.

Although  these devices were designed as test structures, their performance provides important critical insights. The Ta-based sensor exhibits a softer magnetic response; given its domain structure, further miniaturization is expected to promote near  single-domain behavior. Currently, the Ta device provides an optimal linear hysteresis-free range of $\pm 200\text{ }\mu\text{T}$. Although Ta’s high resistivity leads to a lower maximum resistance change (as indicated by the coefficient $A$, Eq.\ref{Eq:Rcos2Params}), it simultaneously enhances the spin Hall effect, thereby increasing the SMR contribution \cite{Sagasta2016}.
An overview of the test device performance is given by: a) the measurement sensitivity is approximately 0.1~$\Omega/\mathrm{mT}$ (Pt), 0.7~$\Omega/\mathrm{mT}$ (Ta); b) the noise level (N.L.) in $H=0$  is around 0.8~$\mathrm{m\Omega}$ (Pt), 3~$\mathrm{m\Omega}$ (Ta); c) the field error is given by $\mu_0 \Delta H$= N.L./sensitivity, resulting in: 8~$\mathrm{\mu T}$ (Pt), 4.2~$\mathrm{ \mu T}$ (Ta).


\subsection{Power Consumption and Optimization}

Apart from high sensitivity and low noise ratio, the reduced power consumption is a primary goal of spintronic sensors. For an SMR-based Wheatstone bridge with four branches, the total power $P$ is:

\beq
P = 4 R_\mathrm{||} I^2 = 4 \frac{R_\mathrm{HM} R_\mathrm{FM}}{R_\mathrm{HM}+R_\mathrm{FM}} I^2
\eeq

When utilizing SOT for linearization, the minimum operating current is dictated by the required linear field range \cite{Xu-SOT_2019}. To achieve linearity within $\pm 0.1\text{ mT}$, the current-induced field $H_I$ (comprising the field-like torque and the Oersted field) must match that magnitude:

\beq
H_I = H_{\mathrm{FL}}+H_{\mathrm{Oe}}=-\frac{[(\hslash/2)/e] j_\mathrm{HM} }{\mu_0 M_s d_\mathrm{FM}} \xi_\mathrm{FL} + j_\mathrm{HM} \frac{d_\mathrm{HM}}{2} \left(1 - \frac{R_\mathrm{HM}}{R_\mathrm{FM}} \right)
\label{EQ:Shift1}
\eeq


This equation defines the minimum current density in the HM layer $j_{\mathrm{HM}}$ necessary to obtain the desired torque. Assuming that $R_{FM} \gg R_{HM}$ (so $I \approx I_{HM}$) gives the minimum power consumption (in presence of a conductive FM layer the power for obtaining the required torque will be always higher):
\beq
P = 4 R_{\mathrm{HM}} w^2 d_\mathrm{HM}^2 H_\mathrm{I}^2 \left[-\frac{[(\hslash/2)/e] \xi_\mathrm{FL}}{\mu_0 M_\mathrm{s} d_\mathrm{FM}} + \frac{d_\mathrm{HM}}{2}\right]^{-2}
\eeq
where $w$ the width of the bridge branch. 

For our Pt-based test sample, we find from the Fuchs-Sondheimer analysis compared with resistivity measurements that for one Wheatstone bridge branch $R_{\mathrm{Pt}}$=390 $\ohm$ (5nm Pt), $R_{\mathrm{FeCoB}}$= 2850 $\ohm$ (2nm FeCoB), while $\xi_\mathrm{FL}$=0.023 was obtained from measurements on Hall bar devices. We can then estimate the minimum power consumption to be approximately 290 mW, which is comparable with the values obtained Ref.\cite{Xu-SOT_2019} and other magnetoresistive sensors \cite{Zheng_sensors_2019}. For Ta, instead, we have $R_{Ta}$= 1730 $\ohm$ (5nm Ta), $\xi_\mathrm{FL}$=-0.01 and the minimum power is 2.17 W. We note here, that the Ta sensor is interesting from the point of view of the magnetic properties while it is less suited considering the electrical properties, having a resistivity comparable to the ferromagnetic layer.

From this equation we derive several optimization rules: a) Maximize the Spin Hall torque efficiency: Increasing $\xi_\mathrm{FL}$ directly reduces the power required to generate the biasing SOT; b) Resistivity Ratio: Power is minimized when $\rho_{\text{HM}} \ll \rho_{\text{FM}}$, which also maximizes the magnetoresistance signal (see section \ref{sec:AMR}). This can be obtained also by choosing a thin FM layer with respect to the HM layer, overall thinner layers reduce the power.  

However, drawing definitive conclusions regarding device performance is remarkably difficult, given the complex relationships between the material parameters. 
These include the electrical resistivities of the HM and FM layers, the saturation magnetization, the effective magnetic anisotropy, and the SOT efficiencies, all of which converge to determine the sensor's response. 
Moreover, the specific geometric configuration—defined by the aspect ratio and dimensions of the bridge branches—further complicates the prediction of the observed magnetoresistance. 
While certain qualitative trends can be identified through observation, locating the precise "sweet spot" within this multi-dimensional parameter space necessitates a complete multi-physics model.



\FloatBarrier
\section{Conclusions}

We developed a comprehensive model based on a multiphysics framework for describing the output of magnetic field sensors based on Spin Hall Magnetoresistance (SMR) in a Wheatstone bridge configuration. By  combining a modified Stoner-Wohlfarth with a Fuchs-Sondheimer analysis of current distribution, we accounted for the complex interplay between SMR, Anisotropic Magnetoresistance (AMR), and Spin-Orbit Torque (SOT). In particular, we included the influence of domain wall motion during reversal. The "truncated astroid" approach effectively reproduces  the hysteresis curves of test devices with non-uniform magnetization. The model parameter $T$ could be linked to the experimental DW density. The model was validated experimentally on Pt/FeCoB and Ta/FeCoB bilayers and showed good agreement. The Ta-based samples, being magnetically softer, offer improved  linearity, while the Pt samples, with lower resistivity, better meet the electrical requirements for well performing sensors. 

Our model demonstrates  that optimizing next-generation spintronic sensors requires a careful  balance between magnetic and electrical properties. Specifically, maximizing the spin Hall torque efficiency, here the field-like torque, reduces power consumption and enhances sensitivity. The model serves  as a tool for eventually finding the "sweet spot" parameters necessary to move beyond the performance of traditional magentoresistive sensors. 

While currently the performance is comparable with AMR sensors, we consider the main advantage of SMR sensors to be the simpler device structure and efficient ways for biasing and linearization through the intrinsic SOT.  While AMR sensors require a complex "barber pole" structure, SMR-based sensors inherently maximize the signal when the current is perpendicular to the magnetization, thus reducing hysteresis. 
The simple device structure, being composed essentially of an ultrathin bilayer, enables use in niche applications requiring optical access and semi-transparency, as previously suggested. An ultimate improvement in performance is expected if it is possible to overcome current limitations imposed by the spin Hall effect and to exploit other mechanisms, such as orbital torques, which may eventually be able to increase the field like torque.

\section*{Acknowledgments}

This research has been funded by the Italian Ministry of University and Research (MUR), “NEXT GENERATION METROLOGY”, FOE 2023 (Ministry Decree n. 789/2023), and by the Italian Ministry of Foreign Affairs and International Cooperation, grant number IN25GR02. Microfabrication was performed at the Academic Centre for Materials and Nanotechnology of AGH. P.W. acknowledges the Excellence initiative-research university (IDUB) programme of the AGH University of Krakow. W.S. acknowledges National Science Centre, Poland project no. 2021/40/Q/ST5/00209 (Sheng). M.K. thanks C.Rinaldi and R.Bertacco for support and fruitful discussions.


\bibliographystyle{elsarticle-num} 
\bibliography{SMR}

@book{Bertotti-1998,
	Author = {G. Bertotti},
	Title = {Hysteresis in Magnetism},
	Publisher = {Academic Press},
	Address = {New York},
	Year = {1998}}

@book{Bertotti-2009,
	Author = {G. Bertotti and I. D. Mayergoyz and C. Serpico},
	Title = {Nonlinear magnetization dynamics in nanosystems},
	Publisher = {Elsevier},
	Address = {Amsterdam},
	Year = {2009}}

@article{Althammer-2013,
  title={Quantitative study of the spin Hall magnetoresistance in ferromagnetic insulator/normal metal hybrids},
  author={Althammer, Matthias and Meyer, Sibylle and Nakayama, Hiroyasu and Schreier, Michael and Altmannshofer, Stephan and Weiler, Mathias and Huebl, Hans and Gepr{\"a}gs, Stephan and Opel, Matthias and Gross, Rudolf and others},
  journal={Physical Review B},
  volume={87},
  number={22},
  pages={224401},
  year={2013},
  publisher={APS}
}

@article{Chen-2013,
  title = {Theory of spin Hall magnetoresistance},
  author = {Chen, Yan-Ting and Takahashi, Saburo and Nakayama, Hiroyasu and Althammer, Matthias and Goennenwein, Sebastian T. B. and Saitoh, Eiji and Bauer, Gerrit E. W.},
  journal = {Phys. Rev. B},
  volume = {87},
  issue = {14},
  pages = {144411},
  numpages = {9},
  year = {2013},
  month = {Apr},
  publisher = {American Physical Society},
  doi = {10.1103/PhysRevB.87.144411},
}

@article{Choi-2017,
  title={Spin Hall magnetoresistance in heavy-metal/metallic-ferromagnet multilayer structures},
  author={Choi, Jong-Guk and Lee, Jae Wook and Park, Byong-Guk},
  journal={Physical Review B},
  volume={96},
  number={17},
  pages={174412},
  year={2017},
  publisher={APS}
}

@article{Dieny-2020,
  title={Opportunities and challenges for spintronics in the microelectronics industry},
  author={Dieny, Bernard and Prejbeanu, Ioan Lucian and Garello, Kevin and Gambardella, Pietro and Freitas, Paulo and Lehndorff, Ronald and Raberg, Wolfgang and Ebels, Ursula and Demokritov, Sergej O and Akerman, Johan and others},
  journal={Nature Electronics},
  volume={3},
  number={8},
  pages={446--459},
  year={2020},
  publisher={Nature Publishing Group}
}

@article{Haney-2013,
  title={Current induced torques and interfacial spin-orbit coupling: Semiclassical modeling},
  author={Haney, Paul M and Lee, Hyun-Woo and Lee, Kyung-Jin and Manchon, Aur{\'e}lien and Stiles, Mark D},
  journal={Physical Review B},
  volume={87},
  number={17},
  pages={174411},
  year={2013},
  publisher={APS}
}

@article{Liu-2011,
  title={Spin-torque ferromagnetic resonance induced by the spin Hall effect},
  author={Liu, Luqiao and Moriyama, Takahiro and Ralph, DC and Buhrman, RA},
  journal={Physical review letters},
  volume={106},
  number={3},
  pages={036601},
  year={2011},
  publisher={APS}
}

@article{Liu-2020,
  title={Determination of Spin-Orbit-Torque Efficiencies in Heterostructures with In-Plane Magnetic Anisotropy},
  author={Liu, Yan-Ting and Chen, Tian-Yue and Lo, Tzu-Hsiang and Tsai, Tsung-Yu and Yang, Shan-Yi and Chang, Yao-Jen and Wei, Jeng-Hua and Pai, Chi-Feng},
  journal={Physical Review Applied},
  volume={13},
  number={4},
  pages={044032},
  year={2020},
  publisher={APS}
}

@article{Miron-2011,
  title={Perpendicular switching of a single ferromagnetic layer induced by in-plane current injection},
  author={Miron, Ioan Mihai and Garello, Kevin and Gaudin, Gilles and Zermatten, Pierre-Jean and Costache, Marius V and Auffret, St{\'e}phane and Bandiera, S{\'e}bastien and Rodmacq, Bernard and Schuhl, Alain and Gambardella, Pietro},
  journal={Nature},
  volume={476},
  number={7359},
  pages={189--193},
  year={2011},
  publisher={Nature Publishing Group}
}

@article{Ripka-2010,
  title={Advances in Magnetic Field Sensors},
  author={Ripka, Pavel and Janosek, Michal},
  journal={IEEE SENSORS JOURNAL},
  volume={10},
  pages={1108},
  year={2010},
  publisher={IEEE}
}

@article{Weiler-2012,
  title={Local charge and spin currents in magnetothermal landscapes},
  author={Weiler, Mathias and Althammer, Matthias and Czeschka, Franz D and Huebl, Hans and Wagner, Martin S and Opel, Matthias and Imort, Inga-Mareen and Reiss, G{\"u}nter and Thomas, Andy and Gross, Rudolf and others},
  journal={Physical review letters},
  volume={108},
  number={10},
  pages={106602},
  year={2012},
  publisher={APS}
}

@article{avci2014,
  title = {Interplay of spin-orbit torque and thermoelectric effects in ferromagnet/normal-metal bilayers},
  author = {Avci, Can Onur and Garello, Kevin and Gabureac, Mihai and Ghosh, Abhijit and Fuhrer, Andreas and Alvarado, Santos F. and Gambardella, Pietro},
  journal = {Phys. Rev. B},
  volume = {90},
  issue = {22},
  pages = {224427},
  numpages = {11},
  year = {2014},
  month = {Dec},
  publisher = {American Physical Society},
}

@article{yihong2021,
  title = {Charge–spin interconversion and its applications in magnetic sensing},
  author = {Yihong Wu and Yanjun Xu and Ziyan Luo and Yumeng Yang and Hang Xie and Qi Zhang and Xinhai Zhang},
  journal = {J. Appl. Phys.},
  volume = {129},
  pages = {060902},
  year = {2021},
  }

@ARTICLE{Magni2022,
  author={Magni, Alessandro and Basso, Vittorio and Sola, Alessandro and Soares, Gabriel and Meggiato, Nicola and Kuepferling, Michaela and Skowroński, Witold and Łazarski, Stanisław and Grochot, Krzysztof and Khanjani, Mehran Vafaee and Langer, Juergen and Ocker, Berthold},
  journal={IEEE Transactions on Magnetics}, 
  title={Spin Hall Magnetoresistance and Spin–Orbit Torque Efficiency in Pt/FeCoB Bilayers}, 
  year={2022},
  volume={58},
  number={2},
  pages={1-5},
  doi={10.1109/TMAG.2021.3084866}}

@ARTICLE{Basso2022,
  author={Basso, Vittorio and Magni, Alessandro and Sola, Alessandro and Kuepferling, Michaela},
  journal={IEEE Transactions on Magnetics}, 
  title={Spin Currents at the Interface and Spin Hall Torque}, 
  year={2022},
  volume={58},
  number={8},
  pages={1-5},
  doi={10.1109/TMAG.2022.3142885}}

@ARTICLE{Demirci2020,
  author={E.Demirci},
  journal={Journal of Superconductivity and Novel Magnetism}, 
  title={Magnetic and Magnetotransport Properties of Memory Sensors Based on Anisotropic Magnetoresistance}, 
  year={2020},
  volume={33},
  pages={3835-3840},
  doi={10.1007/s10948-020-05646-4}}

@article{Xu2017,
    author = {Xu, Yanjun and Yang, Yumeng and Luo, Ziyan and Xu, Baoxi and Wu, Yihong},
    title = {Macro-spin modeling and experimental study of spin-orbit torque biased magnetic sensors},
    journal = {Journal of Applied Physics},
    volume = {122},
    number = {19},
    pages = {193904},
    year = {2017},
    month = {11},
    issn = {0021-8979},
    doi = {10.1063/1.4994109},
}

@article{Yang2017,
    author = {Yang, Yumeng and Xu, Yanjun and Xie, Hang and Xu, Baoxi and Wu, Yihong},
    title = {Semitransparent anisotropic and spin Hall magnetoresistance sensor enabled by spin-orbit torque biasing},
    journal = {Applied Physics Letters},
    volume = {111},
    number = {3},
    pages = {032402},
    year = {2017},
    month = {07},
    issn = {0003-6951},
    doi = {10.1063/1.4993899},
}

@article{Xu2018,
author = {Xu, Yanjun and Yang, Yumeng and Zhang, Mengzhen and Luo, Ziyan and Wu, Yihong},
title = {Ultrathin All-in-One Spin Hall Magnetic Sensor with Built-In AC Excitation Enabled by Spin Current},
journal = {Advanced Materials Technologies},
volume = {3},
number = {8},
pages = {1800073},
doi = {https://doi.org/10.1002/admt.201800073},
year = {2018}
}

@book{Hubert-1998,
	Author = {Alex Hubert and Rudolf Sch\"afer},
	Title = {Magnetic Domains},
	Publisher = {Springer},
	Address = {Springer-Verlag Berlin Heidelberg 1998},
	Year = {1998}}

@article{karlqvist-1954,
  title={Calculation of the magnetic field in the ferromagnetic layer of a magnetic drum},
  journal={Acta polytechnica},
  volume={161},
  author={Karlqvist, Olle},
  year={1954},
  publisher={Elanders boktr.;[H. Lindst{\aa}hls bokhandel i distribution]}
}

@article{Appino-2000,
title = {A vector hysteresis model including domain wall motion and coherent rotation},
journal = {Physica B: Condensed Matter},
volume = {275},
number = {1},
pages = {103-106},
year = {2000},
issn = {0921-4526},
doi = {https://doi.org/10.1016/S0921-4526(99)00709-7},
author = {C Appino and M Valsania and V Basso},
}

@ARTICLE{Mott-1936,
  author={Mott, Nevill Francis},
  journal={Proc. R. Soc. Lond.}, 
  title={The electrical conductivity of transition metals}, 
  year={1936},
  volume={153},
  number={880},
  pages={699-717},
  doi={10.1098/rspa.1936.0031}}

@ARTICLE{McGuire-1975,
  author={McGuire, T. and Potter, R.},
  journal={IEEE Transactions on Magnetics}, 
  title={Anisotropic magnetoresistance in ferromagnetic 3d alloys}, 
  year={1975},
  volume={11},
  number={4},
  pages={1018-1038},
  doi={10.1109/TMAG.1975.1058782}}

@article{Marsocci-1965,
  title = {Effect of Spin-Orbit Interaction on the Magnetoresistance of Single-Crystal Nickel and Nickel-Iron Thin Films},
  author = {Marsocci, Velio A.},
  journal = {Phys. Rev.},
  volume = {137},
  issue = {6A},
  pages = {A1842--A1846},
  numpages = {0},
  year = {1965},
  month = {Mar},
  publisher = {American Physical Society},
  doi = {10.1103/PhysRev.137.A1842},
}

@article{Smit-1951,
title = {Magnetoresistance of ferromagnetic metals and alloys at low temperatures},
journal = {Physica},
volume = {17},
number = {6},
pages = {612-627},
year = {1951},
issn = {0031-8914},
doi = {https://doi.org/10.1016/0031-8914(51)90117-6},
author = {J Smit}}

@article{Appino-2023,
    author = {Appino, C.},
    title = {Exact formulation for hysteresis loops and energy loss in Stoner–Wohlfarth systems},
    journal = {AIP Advances},
    volume = {13},
    number = {5},
    pages = {055018},
    year = {2023},
    month = {05},
    issn = {2158-3226},
    doi = {10.1063/5.0143905},
}

@Article{Kanno2022,
author={Kanno, Akitake
and Nakasato, Nobukazu
and Oogane, Mikihiko
and Fujiwara, Kosuke
and Nakano, Takafumi
and Arimoto, Tadashi
and Matsuzaki, Hitoshi
and Ando, Yasuo},
title={Scalp attached tangential magnetoencephalography using tunnel magneto-resistive sensors},
journal={Scientific Reports},
year={2022},
month={Apr},
day={12},
volume={12},
number={1},
pages={6106},
issn={2045-2322},
doi={10.1038/s41598-022-10155-6},
}

@article{Nguyen2021,
author = {Nguyen, Minh Hai and Pai, Chi Feng},
doi = {10.1063/5.0041123},
issn = {2166532X},
journal = {APL Materials},
number = {3},
pages = {1--15},
publisher = {AIP Publishing, LLC},
title = {{Spin-orbit torque characterization in a nutshell}},
volume = {9},
year = {2021}
}

@article{An2025,
author = {An, Zhengang and Zhang, Lei and Fan, Yanyun and Li, Qingtong and Li, Dachao},
doi = {10.1016/j.sna.2024.116174},
issn = {09244247},
journal = {Sensors and Actuators A: Physical},
number = {December 2024},
pages = {116174},
publisher = {Elsevier B.V.},
title = {{A comprehensive review of TMR current sensors for smart grids: Materials, optimization methods, and applications}},
volume = {382},
year = {2025}
}

@article{barkhausen1919,
  author          = {Barkhausen, Heinrich},
  journal         = {Physikalische Zeitschrift},
  volume          = {20},
  pages           = {401--403},
  title           = {Zwei neue Erscheinungen},
  year            = {1919}
}

@article{renaudin2010,
author = {Renaudin, Valérie and Afzal, Muhammad Haris and Lachapelle, Gérard},
title = {Complete Triaxis Magnetometer Calibration in the Magnetic Domain},
journal = {Journal of Sensors},
volume = {2010},
number = {1},
pages = {967245},
doi = {https://doi.org/10.1155/2010/967245},
year = {2010}
}

@article{Rohrmann2018,
author = {Rohrmann, Kris and Meier, Phil and Sandner, Marvin and Prochaska, Marcus},
doi = {10.1109/ICSENS.2018.8589677},
isbn = {9781538647073},
issn = {21689229},
journal = {Proceedings of IEEE Sensors},
pages = {1--4},
publisher = {IEEE},
title = {{A Novel Methodology for Stray Field Insensitive xMR Angular Position Sensors}},
volume = {2018-October},
year = {2018}
}

@article{Manchon2015,
archivePrefix = {arXiv},
arxivId = {1507.02408},
author = {Manchon, A. and Koo, H. C. and Nitta, J. and Frolov, S. M. and Duine, R. A.},
doi = {10.1038/nmat4360},
eprint = {1507.02408},
issn = {14764660},
journal = {Nature Materials},
number = {9},
pages = {871--882},
pmid = {26288976},
publisher = {Nature Publishing Group},
title = {{New perspectives for Rashba spin-orbit coupling}},
volume = {14},
year = {2015}
}

@article{Sagasta2016,
author = {Sagasta, Edurne and Omori, Yasutomo and Isasa, Miren and Gradhand, Martin and Hueso, Luis E. and Niimi, Yasuhiro and Otani, Yoshichika and Casanova, F{\`{e}}lix},
doi = {10.1103/PhysRevB.94.060412},
issn = {24699969},
journal = {Physical Review B},
number = {6},
pages = {1--6},
title = {{Tuning the spin Hall effect of Pt from the moderately dirty to the superclean regime}},
volume = {94},
year = {2016}
}

@incollection{dibbern1989,
  author       = {Dibbern, U.},
  title        = {Magnetoresistive Sensors},
  booktitle    = {Sensors -- A Comprehensive Survey},
  series       = {Magnetic Sensors},
  volume       = {5},
  editor       = {G{\"o}pel, W. and Hesse, J. and Zemel, J. N. and Boll, R. and Overshott, K. J.},
  publisher    = {VCH},
  year         = {1989},
  pages        = {342--379},
  address      = {Weinheim, Germany}
}

@article{Hoffmann2013,
author = {Hoffmann, Axel},
doi = {10.1109/TMAG.2013.2262947},
issn = {00189464},
journal = {IEEE Transactions on Magnetics},
number = {10},
pages = {5172--5193},
publisher = {IEEE},
title = {{Spin hall effects in metals}},
volume = {49},
year = {2013}
}

@misc{suess2021,
  author       = {Dieter Suess, Udo Ausserlechner, Armin Satz},
  title        = {DEVICE AND METHOD FOR DETECTING A MAGNETIC FIELD USING THE SPIN ORBIT TORQUE EFFECT},
  year         = {2021},
  month        = {October},
  day          = {7},
  number       = {US 2021/0311139 A1},
  type         = {Patent},
  nationality  = {United States},
}

@article{Freitas2016,
author = {Freitas, Paulo P. and Ferreira, Ricardo and Cardoso, Susana},
doi = {10.1109/JPROC.2016.2578303},
issn = {15582256},
journal = {Proceedings of the IEEE},
number = {10},
pages = {1894--1918},
publisher = {IEEE},
title = {{Spintronic Sensors}},
volume = {104},
year = {2016}
}

@article{Chopin-sensor-2020,
author = {Chopin, Chlo{\'e} and Torrejon, Jacob and Solignac, Aur{\'e}lie and Fermon, Claude and Jendritza, Patrick and Fries, Pascal and Pannetier-Lecoeur, Myriam},
title = {Magnetoresistive Sensor in Two-Dimension on a 25 um Thick Silicon Substrate for In Vivo Neuronal Measurements},
journal = {ACS Sensors},
volume = {5},
number = {11},
pages = {3493-3500},
year = {2020},
doi = {10.1021/acssensors.0c01578},
}

@book{Tumanski2011,
  author    = {Sławomir Tumanski},
  title     = {Handbook of Magnetic Measurements},
  edition   = {1st},
  year      = {2011},
  publisher = {CRC Press},
  address   = {Boca Raton},
  doi       = {10.1201/b10979},
  pages     = {404},
}

@ARTICLE{Zheng_sensors_2019,
  author={Zheng, Chao and Zhu, Ke and Cardoso de Freitas, Susana and Chang, Jen-Yuan and Davies, Joseph E. and Eames, Peter and Freitas, Paulo P. and Kazakova, Olga and Kim, CheolGi and Leung, Chi-Wah and Liou, Sy-Hwang and Ognev, Alexey and Piramanayagam, S. N. and Ripka, Pavel and Samardak, Alexander and Shin, Kwang-Ho and Tong, Shi-Yuan and Tung, Mean-Jue and Wang, Shan X. and Xue, Songsheng and Yin, Xiaolu and Pong, Philip W. T.},
  journal={IEEE Transactions on Magnetics}, 
  title={Magnetoresistive Sensor Development Roadmap (Non-Recording Applications)}, 
  year={2019},
  volume={55},
  number={4},
  pages={1-30},
  doi={10.1109/TMAG.2019.2896036}}

@article{Oogane_2021,
doi = {10.35848/1882-0786/ac3809},
year = {2021},
month = {nov},
publisher = {IOP Publishing},
volume = {14},
number = {12},
pages = {123002},
author = {Oogane, Mikihiko and Fujiwara, Kosuke and Kanno, Akitake and Nakano, Takafumi and Wagatsuma, Hiroshi and Arimoto, Tadashi and Mizukami, Shigemi and Kumagai, Seiji and Matsuzaki, Hitoshi and Nakasato, Nobukazu and Ando, Yasuo},
title = {Sub-pT magnetic field detection by tunnel magneto-resistive sensors},
journal = {Applied Physics Express}
}

@article{Xu-SOT_2019,
    author = {Xu, Yanjun and Yang, Yumeng and Xie, Hang and Wu, Yihong},
    title = {Spin Hall magnetoresistance sensor using AuxPt1-x as the spin-orbit torque biasing layer},
    journal = {Applied Physics Letters},
    volume = {115},
    number = {18},
    pages = {182406},
    year = {2019},
    month = {10},
    issn = {0003-6951},
    doi = {10.1063/1.5127838},
}

@article{Korlatan_SOT_2023,
  title = {Single-device offset-free magnetic field sensing with tunable sensitivity and linear range based on spin-orbit torques},
  author = {Koraltan, Sabri and Schmitt, Christin and Bruckner, Florian and Abert, Claas and Pr\"ugl, Klemens and Kirsch, Michael and Gupta, Rahul and Zeilinger, Sebastian and Salazar-Mej\'{\i}a, Joshua M. and Agrawal, Milan and G\"uttinger, Johannes and Satz, Armin and Jakob, Gerhard and Kl\"aui, Mathias and Suess, Dieter},
  journal = {Phys. Rev. Appl.},
  volume = {20},
  issue = {4},
  pages = {044079},
  numpages = {15},
  year = {2023},
  month = {Oct},
  publisher = {American Physical Society},
  doi = {10.1103/PhysRevApplied.20.044079},
}

@article{khan2021magnetic,
  title={Magnetic sensors-A review and recent technologies},
  author={Khan, Mohammed Asadullah and Sun, Jian and Li, Bodong and Przybysz, Alexander and Kosel, J{\"u}rgen},
  journal={Engineering Research Express},
  volume={3},
  number={2},
  pages={022005},
  year={2021},
  publisher={IOP Publishing}
}

@article{li2021spin,
  title={A spin--orbit torque device for sensing three-dimensional magnetic fields},
  author={Li, Ruofan and Zhang, Shuai and Luo, Shijiang and Guo, Zhe and Xu, Yan and Ouyang, Jun and Song, Min and Zou, Qiming and Xi, Li and Yang, Xiaofei and others},
  journal={Nature Electronics},
  volume={4},
  number={3},
  pages={179--184},
  year={2021},
  publisher={Nature Publishing Group UK London}
}

@article{leitao2024enhanced,
  title={Enhanced performance and functionality in spintronic sensors},
  author={Leitao, Diana C and Riel, Floris JF van and Rasly, Mahmoud and Araujo, Pedro DR and Salvador, Maria and Paz, Elvira and Koopmans, Bert},
  journal={npj Spintronics},
  volume={2},
  number={1},
  pages={54},
  year={2024},
  publisher={Nature Publishing Group UK London}
}

@inproceedings{fuchs1938conductivity,
  title={The conductivity of thin metallic films according to the electron theory of metals},
  author={Fuchs, K},
  booktitle={Mathematical Proceedings of the Cambridge Philosophical Society},
  volume={34},
  pages={100--108},
  year={1938},
  organization={Cambridge University Press}
}

@article{cecot_influence_2017,
	title = {Influence of intermixing at the {Ta}/{CoFeB} interface on spin {Hall} angle in {Ta}/{CoFeB}/{MgO} heterostructures},
	volume = {7},
	issn = {2045-2322},
	doi = {10.1038/s41598-017-00994-z},
	language = {en},
	number = {1},
	journal = {Scientific Reports},
	author = {Cecot, Monika and Karwacki, \L{}ukasz and Skowro\'{n}ski, Witold and Kanak, Jaros\l{}aw and Wrona, Jerzy and \.Z{}ywczak, Antoni and Yao, Lide and van Dijken, Sebastiaan and Barna\'{s}, J\'{o}zef and Stobiecki, Tomasz},
	year = {2017},
}

@article{Byeonghwa_2022,
    title = {Advances and key technologies in magnetoresistive sensors with high thermal stabilities and low field detectivities},
    author = {Lim, Byeonghwa and Mahfoud, Mohamed and T. Das, Proloy and Jeon, Taehyeong and Jeon, Changyeop and Kim, Mijin and Nguyen, Trung-Kien and Tran, Quang-Hung and Terki, Ferial and Kim, CheolGi},
	volume = {10},
	doi = {10.1063/5.0087311},
	number = {5},
	journal = {APL Materials},	
	year = {2022},
}

@article{Miron_2010,
  title = {Current-driven spin torque induced by the Rashba effect in a ferromagnetic metal layer},
  author = {Mihai Miron, Ioan and Gaudin, Gilles and Auffret, Stéphane and  Rodmacq, Bernard and Schuhl, Alain and Pizzini, Stefania and Vogel, Jan and Gambardella, Pietro},
  journal = {Nature Materials},
  volume = {9},
  issue = {3},
  pages = {230--234},
  year = {2010},
  doi = {10.1038/nmat2613},
}

@article{Silva_2015,
author = {Silva, A. and Leitao, Diana and Valadeiro, Joao and Amaral, José and Freitas, Paulo and Cardoso, Susana},
year = {2015},
month = {09},
pages = {},
title = {Linearization strategies for high sensitivity magnetoresistive sensors},
volume = {72},
journal = {The European Physical Journal Applied Physics},
doi = {10.1051/epjap/2015150214}
}

@ARTICLE{skowronski-2021,
  author={Skowroński, Witold and Grochot, Krzysztof and Rzeszut, Piotr and Łazarski, Stanisław and Gajoch, Grzegorz and Worek, Cezary and Kanak, Jarosław and Stobiecki, Tomasz and Langer, Jürgen and Ocker, Berthold and Vafaee, Mehran},
  journal={IEEE Transactions on Electron Devices}, 
  title={Angular Harmonic Hall Voltage and Magnetoresistance Measurements of Pt/FeCoB and Pt-Ti/FeCoB Bilayers for Spin Hall Conductivity Determination}, 
  year={2021},
  volume={68},
  number={12},
  pages={6379-6385},
  doi={10.1109/TED.2021.3122999}}

\end{document}